
\input amssym.def
\input amssym.tex

\input psbox.tex
\psfordvips

\def\bbbc{{\mathchoice {\setbox0=\hbox{$\displaystyle\rm C$}\hbox{\hbox
to0pt{\kern0.4\wd0\vrule height0.9\ht0\hss}\box0}}
{\setbox0=\hbox{$\textstyle\rm C$}\hbox{\hbox
to0pt{\kern0.4\wd0\vrule height0.9\ht0\hss}\box0}}
{\setbox0=\hbox{$\scriptstyle\rm C$}\hbox{\hbox
to0pt{\kern0.4\wd0\vrule height0.9\ht0\hss}\box0}}
{\setbox0=\hbox{$\scriptscriptstyle\rm C$}\hbox{\hbox
to0pt{\kern0.4\wd0\vrule height0.9\ht0\hss}\box0}}}}

\def\cc#1{\hfill\quad#1\ \ \hfill}
\def\tv{\tvi\vrule}
\def\tvi{\vrule height 8pt depth 3pt width 0pt}

\def\opl{\mathop{\oplus}\limits}

\def\build#1_#2^#3{\mathrel{\mathop{\kern 0pt#1}\limits_{#2}\limits^{#3}}}
\def\dyad#1#2{\mid #1\rangle\langle#2\mid }
\def\ket#1{{}\mid #1\rangle{}}

\def\obraket#1#2#3{{}\langle#1\mid #2\mid #3\rangle{}} 

\def\bbbr{{\rm I\!R}}

\def\cqfd{\vrule height 0.7 em depth 0.2 em width 0.35 em}

\font\eightrm=cmr8

\font\grand=cmr10 scaled\magstep2

\magnification=\magstep1

\null
\vskip 3cm
\centerline {\bf\grand On hybrid states of two and three level atoms}

\vskip 2cm

\centerline{Claude Billionnet}
\centerline{ Centre de Physique Th{\'e}orique, Ecole Polytechnique, 91128 Palaiseau Cedex, France}
\centerline{Tel: (33)(1) 69 33 47 20}
\centerline{Fax: (33)(1) 69 33 30 08}
\centerline{ E-mail~: billionnet@cpht.polytechnique.fr}

\vskip 2cm

{\bf Abstract}
We calculate atom-photon resonances in the
Wigner-Weisskopf model, admitting two photons and choosing a particular coupling function. We also present a rough description of
 the set of
re\-sonances in a model for a three-level atom coupled to a scalar-photon field. We
give a
general picture of matter-field resonances these results fit into.

\bigskip
Subject area: Quantum Field Theory

\vfill\eject

{\bf I. Introduction}

Last years there has been a renewed interest for atom-radiation
states in non-relativis\-tic quantum electrodynamics. After the rigorous
proof$^1$ of the existence, for small coupling constants, of resonances (singularities of an analytic continuation of
the Hamiltonian resolvent), resonances coming from the
naked-atom-Hamiltonian
 eigenvalues, many studies
have been
concentrated on the study of the Hamiltonian spectrum and more
particularly on the existence of the fundamental state in various models$^{2,3,4,5,6,7}$, for arbitrary
coupling constants. This latter question is a non trivial problem since it has long been known that negative Hamiltonian
eigenvalues may appear when the coupling constant increases$^{8,9}$. In
preceding works (see Ref. 10 and references therein), for our part,
we were interested in a method for
calculating resonances for arbitrary values
of the coupling constant. It is a way of catching the above
mentioned particular eigenvalues. The result we obtained can be generally
stated as follows. The coupling of a discrete-level system $S$ to a
zero-mass field does not only
shift the level energies into the complex plane. Certainly these energies
become resonances. But the coupling also creates other poles of the
resolvent (or of its continuations), which have to be placed on the same
footing as the preceding ones. For certain values of the parameters of
$S$, or certain values of the ($S$+{\it field}) coupling constant,
these latter poles may be eigenvalues of the coupled-system
Hamiltonian. (In the paper, the
word resonance will refer to such a pole or, by extension, to an
eigenvalue, when there is an obvious possible continuous transition from the
one to the other.)

The presence of these poles is well-known for an atom or a molecule
in an environment in which the only emitted or absorbed photons are
monochromatic (lasers or cavities). Indeed, for a two-level atom with
energies  $e_0$ and $e_1$, and photons with energy $e_1-e_0$, the upper level is split into two levels by the coupling, when at most one photon is
considered$^9$ (vacuum-field Rabi splitting). The coupling increases the number
of Hamiltonian eigenvalues because it splits the
degeneracy of each of the eigenvalues $e_1,\
2e_1-e_0,\cdots,\ e_1+n(e_1-e_0),\cdots$ of the uncoupled atom-photon system.
This phenomenon occurs in different but analogous situations: for
instance the coupling of an exciton to the mode of a
cavity$^{11,12}$, and it can be detected by spectroscopic means. It
is also present in
electron-phonon interactions$^{13,14}$. Let us note
that the final number of atom-photon states is simply a consequence of
the combination of discrete atom states with discrete photon-states.

Many papers have studied the coupling of a two-level system to another
system which has either discrete levels or a continuum of levels, often
focusing on the continuous transition from the vacuum Rabi
splitting to the Fermi's golden
rule$^{15,16}$. In the present work, a splitting
is exhibited also in the continuum case. The
eigenstates of the naked atom we are used to are thus but a small part of all
the possible states.

In fact, the appearance of new resonances when the system $S$ is coupled to a
field is a general phenomenon whose explanation is
given in Section II. It is, of course, not due to the smallness of the photon-state
width.  A numerical method, explained and illustrated in previous
publications, enabled us to study and calculate these
resonances in various simple models: $S$ was either a harmonic
oscillator$^{17}$ or a two-level atom. But in this latter case, we
almost always limited ourselves to considering physical states with at most one
photon.

A reason to carry on with the study is that the new resonances
appearing with the coupling might play a role for systems
having a large spatial extension: for instance Rydberg atoms or, a case maybe
more important, molecular orbitals of big molecules. Indeed, in a very
simplified model for an interaction of $S$ with the field, a model in which
the spatial extension $\delta$ of $S$ may be varied ($S$ is a charged
harmonic oscillator whose mass and spring constant may be
varied), we noted that some among the poles we speak about come to
the negative real axis when $\delta$ gets large enough$^{18}$. These poles then correspond to stable states and are therefore
important states. Their wave function can be written: it is a mixing
of electron and photon states. Analogous phenomena may be expected
for more complex extended systems.

In this perspective, it is important to be able to start the study of
the resonances in the two following situations: a system with more
than two levels and, in the case of two levels, a system with several
(non monochromatic) photons. Section III deals with this second
question. To the best of our knowledge, it is tackled in the literature only when the fundamental
state question is discussed$^{4}$. Here we calculate
some mixed (or hybrid)
states with several photons, or the resonances which correspond to
them. Section IV tackles the same problem for three-level atoms, but
since the
situation is more complicated, we limit ourselves to a qualitative
description of the numerous resonances. Both studies lead to the
reasonable conjecture that the number of resonances should be the
product of the number of atom states by the number of independent
radiation states actually coupled to the atom.

\medskip
{\bf II. Notation of the resonances}

We need a precise notation for the different poles.

It will follow from a very general argument, which also
explains why every atomic level should give rise to a double infinity
of resonances. Let us consider a material quantum system $S$ whose
Hamiltonian, $H_S$, has eigenstates $\ket 0,\ket 1$, $\ket
2,\cdots$, with energies $e_0,e_1,e_2,\cdots$. Let us suppose that this system is coupled to the field of a
massless boson, here the photon. We denote the state-space of the
($S$ + {\it field})-system by ${\cal E}$.

Let us consider a very general form for the Hamiltonian of the coupled
system:
$$
H(\lambda)=H_S\otimes 1+ 1\otimes H_{\rm rad}+ \lambda\ V   \eqno (2.1)
$$
where $H_{\rm rad}$ is the energy operator for the photon field and $\lambda\ V$  represents
the coupling of $S$ to the field. Let us introduce the auxiliary
Hamiltonian.
$$
H(\lambda,\mu)=H_S\otimes 1+ \mu\ 1\otimes H_{\rm rad}+ \lambda\ V
                                                       \eqno (2.2)
$$
where $\mu$ is a parameter which may be zero or positive.

\noindent
If $\mu=0$ and $\lambda=0$, the energy levels $e_0,e_1,\cdots$ are infinitely degenerated
(in  ${\cal E}$), as the number of photons accompanying state $\ket m$   may be any
integer and, moreover, the dimension of the space of possible photon
states is infinite. Thus the dimension of the eigenspace ${\cal H}_i$ associated with
the eigenvalue $e_i$ of $H(0,0)$ is infinite. Let us underline the fact that the
degeneracy we speak about here is different from the one we mentioned
in the second paragraph of the introduction.

\noindent
The idea underlying our present study, as well as the preceding ones,
is to perturb $H(0,0)$ with respect to $\lambda$ and
$\mu$ successively. A priori, the perturbation with respect to $\lambda$ removes the
degeneracy, although it may be only partially, in some particular
cases. This leads to the following notation.

{\bf Notation 1}
{\it 
$\zeta_{i,1}(\lambda),\ \zeta_{i,2}(\lambda),\cdots$ denote the $I\!\!N^2\
(=I\!\!N)$ eigenvalues of $H(\lambda,0)$ which tend to
$e_i$ when $\lambda$ tends to $0$.}

The choice of the second index, or an equivalent choice, will be
described in the different contexts.

The perturbation with respect to $\mu$ then leads us to the following
notation for the resonances we are interested in:

{\bf Notation 2}
{\it When we have selected, in a way which remains to be defined, one
pole 
among those of the resolvent of
$H(\lambda,\mu)$ which tend to $\zeta_{i,j}(\lambda)$ when $\mu$
tends to $0$, we denote it by
$z_{i,j}(\lambda,\mu)$ . }
\smallskip
\noindent

There are two reasons for having to make a selection. The first one is
that if the $\zeta_{i,j}$'s remain degenerated, and this will be the case in the model
of Section III (see Proposition 3.2), then the degeneracy may be removed
for $\mu\not=0$; in other words, for $\mu\not=0$, there may be several
resonances (or eigenvalues) going to the same $\zeta_{i,j}$  when $\mu$  goes to
$0$. (See a more precise description at the beginning of Section
III.4.2.) The one we call $z_{i,j}(\lambda,\mu)$    will be a particular one, selected in a way
which will be made precise in the context. The second reason is that
the matrix elements of $[z-H(\lambda,\mu)]^{-1}$   are multivalued
functions of $z$ for $\mu\not=0$. Therefore, two poles of two different determinations may have the
same limit when $\mu$  goes to $0$. However, we assume that there is
only one pole tending to $\zeta_{i,j}$    for a given determination, once the first
choice has been made. This will enable us to get the $z_{i,j}(\lambda,\mu)$'s   by numerical
calculus, starting from their germs $\zeta_{i,j}(\lambda)$'s. The notations of the poles of the
different branches of the resolvent matrix-elements will be made
precise later on, by adding an upper index which refers to the
determination (see Section III.4.2.3).

\medskip
{\bf III Hybrid states for a two-level atom in the 
Wigner-Weisskopf model}

Two resonances (distinct or not, see below) are well-known in the Wigner-Weisskopf model, or in the
Friedrichs model, which
describe a two-level atom coupled to the field of a massless scalar
boson$^{4,8,9}$. These resonances can be seen without considering two-photon
states. One can be associated, for small coupling
constants, to the excited state of the atom; the other one is an eigenvalue
corresponding to a stable state which differs from the unperturbed
fundamental state. This eigenvalue appears when the coupling constant
gets large enough. (Up to recently, it was not clear whether this two
resonances were two occurrences of the same resonance or not. In fact
this depends on the parameters of the physical system, but in any case
it can be shown that there are actually two different resonances$^{10}$,
for a given value of the coupling constant). In this section we want to study other resonances by
taking into account several photons. The existence of such states is
already alluded to in Ref. 4, where Theorem 2.2 shows that the
ground state of the coupled system must take several-photon
states into account. In a particular coupling, we will construct two
of these
states or resonances. It is the subject of Section III.4.2. In
accordance with the general notations introduced in Section II, they
will be denoted by $z_{0,2}(\lambda,1)$ and
$z_{1,2}(\lambda,1)$, tending respectively to the energies $0$ and $1$
of the naked atom. The index 2 will be explained later on.    

\vfill\eject                   
{\bf III.1 The model and some notations}

The  atom state-space is $\bbbc^2$. The fundamental state is
$\ket 0:=(1,0)$, with energy $e_0=0$ and the excited state is $\ket
1:=(0,1)$, with energy $e_1=1$. In the $\{\ket{0},\ket{1}\}$ basis,
 the annihilation operator is $a=\pmatrix{0&1\cr0&0\cr}$.
The field state-space is ${\cal
F}:=\build\oplus_{n=0}^{\infty}\  {\cal F}_n$, as usual, with $ {\cal F}_0:={\bf
C}$ and $ {\cal F}_n:=L^2(\bbbr)^{\vee^n}$. Let $g$ be in $ L^2(\bbbr)$ with $||g||_{_2}=1$ and
$c(\overline g)$, $c^*(g)$ be the operators annihilating and creating
a (scalar) photon in state $g$. Let $H_{\rm rad}$ be the energy
operator in ${\cal F}$.

\noindent
The Hamiltonian of the model is
$$
H(\lambda):=a^*a\otimes 1+1\otimes H_{\rm rad}+\lambda\big(a^*\otimes
c(\overline g)+a\otimes c^*(g)\big)                               \eqno (3.1)
$$
In order to have the mixed states appear readily, we introduce
$$
H(\lambda,\mu):=a^*a\otimes 1+\mu\ 1\otimes H_{\rm
rad}+\lambda\big(a^*\otimes c(\overline g)+a\otimes c^*(g)\big)       \eqno (3.2)
$$
Note that $H(\lambda,\mu)$ is unitarily equivalent to the Hamiltonian
obtained through replacing $g$ in (3.1) by the dilated function
$g_{\mu}(p):=\mu^{-{1\over 2}}g(\mu^{-1}p)$. Introducing the parameter
$\mu$ is thus not so arbitrary as it may seem since $\mu$ is in some way
related to the width of the coupling function. But note that the
position of the peak of $g$ is also moved. This is a departure from a
study which would start with monochromatic photons and enlarge the
width of their spectrum little by little, without changing the
position of the peak.

\noindent
This Hamiltonian has invariant subspaces. To describe them and,
in the same way, prepare the notations for Section IV, let us set

$ {\cal H}_{\rm rad}^{(1)}$, the space spanned by $g$

$ {\cal F}^{(1)}_n:= \big({\cal H}_{\rm rad}^{(1)}\big)^{\vee^n}$, the
space of
 $n$ photon states, but each photon being in state $g$.

${\cal E}_0:=\ket 0\otimes {\cal F}_0$, and for $n\geq 1$,
${\cal E}_n:=\ket 0\otimes {\cal F}_n  \oplus \ket 1\otimes {\cal
  F}_{n-1}$,  a space
we call the\break\indent `` $n$-excitation space''.

${\cal E}_n^{(1)}:=\ket 0\otimes {\cal F}_n^{(1)}  \oplus \ket 1\otimes
{\cal F}_{n-1}^{(1)},\quad n\geq 1,\quad  {\cal E}_0^{(1)}={\cal E}_0 $  

${\cal E}^{(1)}:=\opl_0^\infty{\cal E}^{(1)}_n$,\quad ${\cal E}:=\opl_0^\infty{\cal E}_n$

{\bf Lemma} {\it ${\cal E}_n$ is invariant
by $H(\lambda,\mu)$, for all $n\geq 0$   and all $\mu\geq 0$; ${\cal E}_n^{(1)}\subset {\cal E}_n$ is invariant
by $H(\lambda,0)$, for all $n\geq 0$.}

As we said when we introduced Notations 1 and 2 in Section II, the method for
studying the resonances of $H(\lambda,\mu)$ is a numerical method in
which $\mu$ takes greater and greater values, starting from $0$. So we
begin with giving some properties of  $H(\lambda,0)$.

\smallskip
{\bf III.2 Construction of resonances of $H(\lambda,\mu)$ from
eigenvalues of $H(\lambda,0)$. Setting up}

{\bf Proposition 3.1} {\it $H(\lambda,0)\upharpoonright_{{\cal E}^{(1)}} $,
  the restriction of $H(\lambda,0)$ to ${\cal E}^{(1)}$, has a
  double infinity of eigenvalues}
$$
\zeta_{0,n}(\lambda)=2^{-1}(1-\sqrt{1+4n\lambda^2})\ ,\quad
\zeta_{1,n}(\lambda)=2^{-1}(1+\sqrt{1+4n\lambda^2})            \eqno (3.3)
$$
{\it For each $n$, an eigenvector of
  $H(\lambda,0)\upharpoonright_{{\cal E}^{(1)}} $ associated with
$\zeta_{i,n}(\lambda)$ is 
$$
\phi_{i,n}^{(0)}:=\big(1+n\lambda^2\zeta_{0,n}^{-2}(\lambda)\big)^{-1}\Big(\ket{1,g^{\otimes
(n-1)}}+\sqrt
n\ \lambda\  \zeta_{0,n}^{-1}(\lambda)\ket{0,g^{\otimes n}}\Big)\quad\in{\cal
E}_n^{(1)}                          \eqno (3.4)
$$}
\underbar {Proof} $\ {\cal E}_n^{(1)}$, two dimensional, is invariant by
$H(\lambda,0)$.  (3.3) is thus obtained through diagonalizing a two-by-two
matrix. \cqfd

The choice of the first index in the notation of the eigenvalues is
in accordance with the principle stated at the end of Section II. By
the choice of the second index, we indicate that the
eigenvector belongs to  ${\cal E}_n^{(1)}$. 

We note that only some part of the double degeneracy mentioned in Section II
is removed here. Indeed, each $\zeta_{i,n}(\lambda)$ is still
degenerated, since adding an arbitrary number of photons orthogonal to
$g$ to some state does not change the energy of this state. This
follows from [ $H(\lambda,0),1\otimes c^*(h)]=0$, when $(g,h)=0$. Let us
develop this in order to introduce some notations.

Let $g_1,g_2,\cdots$ be a basis of functions orthogonal to $g$. For
$p\geq 1$, let
${\cal G}_{0,p}$ be the subspace of ${\cal E}$ spanned by
$$
\phi_{0,n}^{(n_1,n_2,\cdots)}:=\prod_{i=1}^\infty\ (1\otimes c^*(g_i)\big)^{n_i}\
\phi_{0,p}^{(0)}\quad  \in{\cal E}_{n}                       \eqno (3.5)
$$
when $n$ varies from $p$ to infinity; $n_i$ are $k$
non-negative integers, $k$ being arbitrary and $\displaystyle\sum_{i=1}^k n_i=n-p$.

\noindent
In the same way, let ${\cal G}_{1,p}$ be the subspace spanned by 
$$
\phi_{1,n}^{(n_1,n_2,\cdots)}:=\prod_{i=1}^\infty \ \big(1\otimes c^*(g_i)\big)^{n_i}\
\phi_{1,p}^{(0)}\quad    \in{\cal E}_{n}                \eqno (3.6)
$$
{\bf Proposition 3.2} {\it For generic $\lambda$-values, ${\cal
G}_{i,p}$ is the eigenspace of $H(\lambda,0)$ associated with the
eigenvalue $ \zeta_{i,p}$, for $i=0,1$ and $p\geq 1$.
}

See hints for the proof in Appendix B1, which is devoted to the proof
of an analogous property.

Let us thus note that eigenstates of $H(\lambda,0)$ associated with
$\zeta_{i,n}$ not only are not\break $n$-photon states but are neither 
necessarily $n$-excitation states. This complicates the picture of these
eigenvalues and of the resonances $z_{i,n}(\lambda,\mu)$ they give
rise to.

\noindent
To be complete, let us still mention that $0$ is an eigenvalue of
$H(\lambda,0)$, with, if $\sum n_i=n$, $n$ arbitrary, the associated eigenvectors
 $$
\phi_{0,0}^{(n_1,n_2,\cdots)}:= \prod_{i=1}^\infty\ \big(1\otimes c^*(g_i)\big)^{n_i}\
\ket 0\otimes \Omega\quad    \in{\cal E}_{n}                 \eqno (3.7)
$$
where $\Omega$ denotes the vacuum state of the field.

Let us now turn to the $\mu$-variation, in order to define and construct
resonances 
$z_{0,1}(\lambda,\mu)$, $\ z_{0,2}(\lambda,\mu),..$ and
$z_{1,1}(\lambda,\mu),\ z_{1,2}(\lambda,\mu),\ ..$, from their germs
$\zeta_{i,j}(\lambda)$ (for the notation, see
Notation 2 in Section II). 

\noindent
This construction, as we said, is purely numerical for the moment. It
goes step-by-step, starting from $\mu=0$. We do not look for any
existence theorem nor for a complete description, difficult to get because
of the complicated structure of the set of resonances. We just want to
obtain numerical values for resonances which have not been considered
up to now, and might be important.

\noindent
This requires a particular choice for $g$.

\noindent
Before that, let us nevertheless give some general indications.

\smallskip

{\bf III.3 General remarks about the poles of the resolvent of
$H(\lambda,\mu)$}

In front of so many resonances, it is natural to ask
oneself the question: which one is the one we are used to, that
is to say is there any which could be ``associated'' with the excited
state $\ket 1$, for the Hamiltonian $H(\lambda,1)$? Without going into
details here, let us explain shortly why the issue is not simple. In
particular, given a value $\lambda_0$ of the coupling constant, even a
small one, why
$z_{1,1}(\lambda_0,\mu)$ is not necessarily the resonance we are
 used to. The resonance we are used to is obtained
through restricting $H(\lambda,\mu)$ to ${\cal E}_1$ and following
the resonance which sits at $1$ for $\mu=1$ and $\lambda=0$. We
follow it as $\lambda$ increases from $0$ to $\lambda_0$. Since its
position at $\lambda=0$ does not depend on $\mu$, it amounts to following
the resonance along the path of Figure 1a, from its value $1$ at the
origin of the path.
$$\psboxto (13cm;0cm){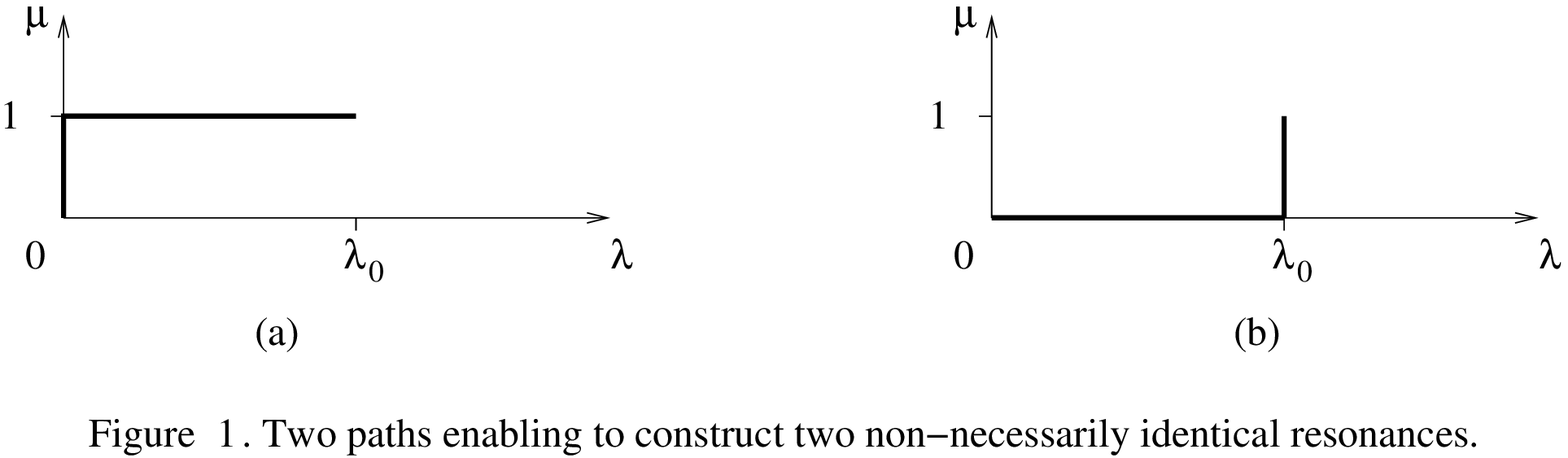}$$
It is the usual perturbative approach, in which there is no need to
introduce $\mu$. On the contrary, $z_{1,1}(\lambda_0,1)$, also defined
with ${\cal E}_1$ (see III.4.1.2), is the limit of
$z_{1,1}(\lambda_0,\mu)$, when $\mu$ increases
from $0$ (then $z_{1,1}(\lambda_0,0)=\zeta_{1,1}(\lambda_0)$) to $1$. Since $\zeta_{1,1}(\lambda_0)$, by definition, tends to $1$
when $\lambda_0$ tends to $0$, $z_{1,1}(\lambda_0,1)$ is the resonance
obtained through following the resonance along the path of Figure 1b,
from the value $1$ at the origin of this path. Now, in some examples, functions
$z_{0,1}(\lambda,\mu)$ and $z_{1,1}(\lambda,\mu)$ are two branches of
a unique analytic 2-variable function, which, even when restricted to
$\bbbr^2$, has branch points. These real branch points may lie inside
the rectangle $[0,\lambda_0]\times [0,1]$ of the $(\lambda,\mu)$ plane,
so that the two previous values of the resonances at the common end
$(\lambda_0,1)$ of the two previous paths may be different. The
$\lambda$ and $\mu$ variations do not necessarily commute$^{10}$. (Let us
note here that it
is thus dangerous to
move $\lambda$ and $\mu$ simultaneously without caution, as we did it
for instance in the second paragraph of
Section III of Ref. 17). This
phenomenon is at the origin of the remark in parentheses in the first
paragraph of Section III.

This seems to be a drawback of the introduction of parameter
$\mu$. But this introduction enables us to see resonances for the
physical Hamiltonian ($\mu=1$) which the perturbative approach does
not give so easily. Indeed, when $\lambda$ is the only parameter and
when it is varied from $\lambda_0$ to $0$, the continuous variation of
the resonance siting at $z_{i,1}(\lambda_0,\mu)$
for $\lambda=\lambda_0$ may lead to $1$, but also to infinity, or to a
pole of $g$ (see Figure 2, where the pole is $-i$). The result
depends on $\mu$ and $g$. But these two latter limits are too singular
points to start a calculation in their neighbourhood.

Since the difficulties of a general study are due to branch
points whose positions depend on $g$, we choose a particular $g$. We now
come back to the study of the announced particular case.

\smallskip

{\bf III.4  Construction of resonances of $H(\lambda,\mu)$ from
eigenvalues of $H(\lambda,0)$. Calculations for a particular $g$.}

We fix the coupling constant at $0.1$ and take the particular function
we already used$^{10}$:
$$
g(p)=\sqrt{2\over\pi}{p\over 1+p^2}                 \eqno (3.8)
$$                      
This is a simple rational function which exhibits the type of
singularity that actual coupling functions may have. (See for instance
matrix elements of the interaction Hamiltonian for hydrogenic atoms in
the electromagnetic field$^{19}$)

\noindent
In Section III.4.1, we recall the definition of $z_{0,1}(\lambda,\mu)$ and
$z_{1,1}(\lambda,\mu)$ and the known formulas through which they are
obtained. We give their values for various $\mu$ and
$\lambda$. We use them later on. Then, in Section III.4.2, we
define resonances $z_{0,2}(\lambda,\mu)$ and $z_{1,2}(\lambda,\mu)$
and give some approximate values. An evaluation of the errors is given
in Appendix A.

\smallskip

{\bf III.4.1 Brief review about two resonances obtained with
only one photon: $z_{0,1}(0.1,\mu)$ and $z_{1,1}(0.1,\mu)$}

They are poles of $\obraket
{1\otimes\Omega}{[z-H\upharpoonright_{{\cal E}_1}]^{-1}}
{1\otimes\Omega}$ or of its analytic continuation.

{\bf III.4.1.1 Resonance $z_{0,1}(0.1,\mu)$}

$z_{i,1}(0.1,\mu)$ are zeros of $f(\lambda,\mu,.)$, defined by   
$$
f(\lambda,\mu,z):=z-1-2\lambda^2\int_0^\infty {g(p)^2\over z-\mu p}dp\eqno (3.9)
$$
in the cut plane $\bbbc\slash\bbbr^+$, or of its
analytic continuation through the cut, clockwise$^{9,20,21}$. For $\mu\leq\mu_c(\lambda):=2\lambda^2\int_0^\infty {g(p)^2\over
p}dp$,
$f(\lambda,\mu,.)$ has only one zero; it is on the negative real axis
and is $0$ if and only if $\mu=\mu_c(\lambda)$. When $\mu$ tends to
$0$, it tends to $\zeta_{0,1}(0.1)$ and therefore we denote it by 
$z_{0,1}(0.1,\mu)$. We recall that the corresponding normed
eigenvector, which we denote by
$\phi_{0,1}^{(0)}(\lambda,\mu)$, is proportional to

$$
\psi_{0,1}^{(0)}(\lambda,\mu):=
\ket{1,\Omega}+\lambda\ket{0,{g\over z_{0,1}(\lambda,\mu)-\mu |p|}} \eqno (3.10)
$$
For
$\mu>\mu_c(\lambda)$, $z_{0,1}(\lambda,.)$ is defined as a zero of the
analytic continuation
$$
f_+(\lambda,\mu,z):=z-1-2\lambda^2\int_0^\infty {g(p)^2\over z-\mu
p}dp+4i\pi{\lambda^2\over\mu}g({z\over\mu})^2                 \eqno (3.11)
$$
of $f(\lambda,\mu,.)$, clockwise through the cut. When $\mu\rightarrow
\mu_c(\lambda)$, it connects to the values for
$\mu<\mu_c(\lambda)$. For $\mu$ varying from $0$ to $2$, some values
of $z_{0,1}(0.1,\mu)$ are given in Table 1 and Figure 1.
$$\vbox{\offinterlineskip\halign{\tv\cc{$#$}\tv&\cc{$\!#$}&\cc{$#$}&\cc{$#$}&\cc{$#$}&\cc{$#$}&\cc{$#$}&\cc{$#$}&\cc{$#$}\tv\cr
\noalign{\hrule}
10^3\ \mu&0&0.1&1&3&6&6.2&6.36&6.366\cr
\noalign{\hrule}
10^4\ z_{0.1}(0.1,\mu)&-99&-94&-68&-34 &-2.5&-1.1&0.04&0.001\cr
\noalign{\hrule}}
}$$
\vskip-0.3cm
\centerline{Table 1. Values of $z_{0,1}(0.1,\mu)$, for some
values of $\mu$ below $\mu_c(0.1)$.}
\vfill\eject
\medskip

$$
\psannotate{\psboxto(0cm;4cm){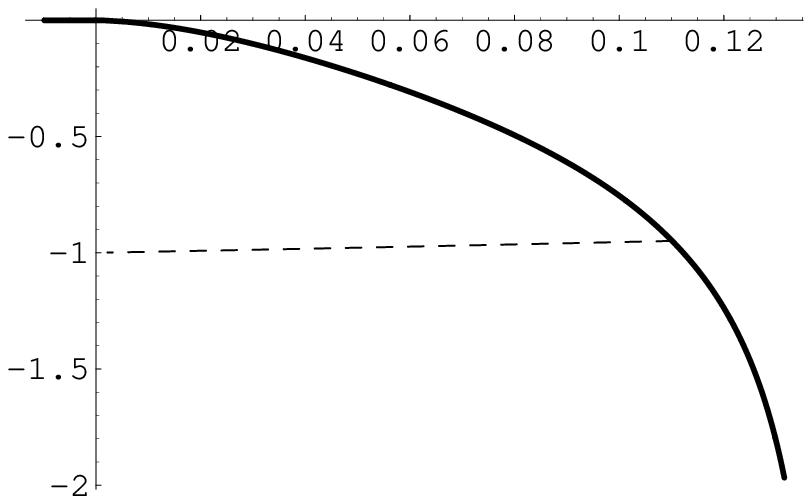}}{\at{-1cm}{3.5cm}{$\mu=0\
^{\nearrow}$}\at{4.1cm}{1.8cm}{$\mu=1\
^{\nearrow}$}\at{5cm}{-0.1cm}{$\mu=2\ ^{\nearrow}$}\at{-2.7cm}{-0.7cm}{Figure
2. The resonance $z_{0,1}(0.1,\mu)$, for $\mu\in
[0,2]$, first in $\bbbr^-$, then in the }\at{-1.2cm}{-1.1cm}{
second sheet of the complex plane}}
$$
\vskip 1cm
The physical value for $\mu=1$ is $0.11-0.95\ i$. For
$\mu=0$, $z_{0,1}(0.1,0)=\zeta_{0,1}(0.1)\simeq
-0.0099$~; the graph goes through $0$ for $\mu\simeq 6.36\  10^{-3}$ and $z_{0,1}(0.1,2)=0.13-1.97\ i$.

In order to connect these results to the usual perturbative treatment,
we drew a dashed line in Figure 2. It describes the continuous move of
the resonance which sits at the point $z_{0,1}(0.1,1)$ for
$\lambda=0.1$, when $\lambda$ decreases from this value to $0$. One
can see that the limit is not $1$, the energy of the naked excited
state, but $-i$, a pole of $g$. This makes a difference with what
occurs for $z_{1,1}(\lambda,1)$, as we will see it just below. To
distinguish the two behaviours, we called a resonance
which, as a function of the only $\lambda$ variable, does not tend to
the excited state energy when $\lambda$ tends to $0$ a ``non standard''
resonance$^{10}$. Be careful that a resonance may be standard for some $\mu$'s
and non standard for others.

{\bf III.4.1.2 Resonance $z_{1,1}(0.1,\mu)$}

It is another zero of $f_+(\lambda,\mu,.)$, which tends to $\zeta_{1,1}(0.1)$, when
$\mu\rightarrow 0$. Its displacement when $\mu$ varies is given by the
full line of Figure 3.
$$
\psannotate{  \psboxto(0cm;4cm){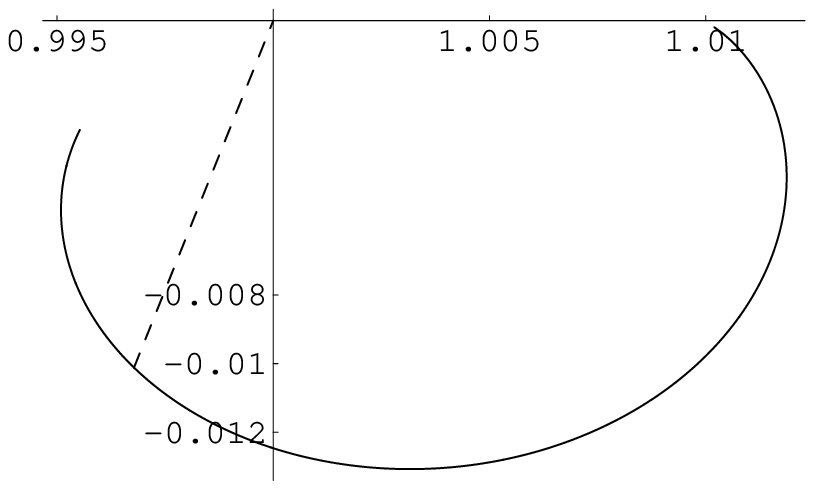}  }  {  \at{6.2cm}{4cm}{$_\swarrow\mu=0$}
\at{-0.1cm}{0.9cm}{$\mu=1\rightarrow$}  \at{-0.6cm}{2.9cm}{$\mu=2\rightarrow$} \at{-2cm}{-0.5cm}{Figure
3. $z_{1,1}(0.1,\mu)$, in the second sheet, for $\mu\in
[0,2]$ (full line)}}
$$
\vskip 0.5cm
\noindent
We have $z_{1,1}(0.1,0)=1.0099\ ,\ z_{1,1}(0.1,1)=0.997-0.010\ i\ ,\
z_{1,1}(0.1,2)=0.995-0.0032\ i$. It can be shown that the
resonance which sits at point $z_{1,1}(\lambda,1)$ for $\lambda=0.1$
moves continuously to $1$ along the dashed line of Figure 3 when $\lambda$ goes
to $0$. Therefore, referring to the discussion of Section III.3, we can
say that $z_{1,1}(0.1,1)$ is the resonance usually associated with the
excited state.

\noindent
We now come to the original part of this Section III.

\smallskip
{\bf III.4.2 Two other resonances, obtained with two photons}

Among the various resonances tending to $\zeta_{i,2}(\lambda)$ when
$\mu$ goes to $0$, we select poles of matrix elements of
$[z-H(\lambda,\mu)\upharpoonright_{{\cal E}_2}]^{-1}$, or of one of
their analytic continuation. Note that another choice among those
mentioned after Notation 2 would have been to look for poles of matrix
elements of $[z-H(\lambda,\mu)\upharpoonright_{{\cal E}_3}]^{-1}$,
since we have eigenvectors of $H(\lambda,0)$ in ${\cal E}_3$
associated with the same eigenvalue $\zeta_{i,2}(\lambda)$: for instance
$\big(1\otimes c^*(g_1)\big)\phi_{i,2}$. But we make no claim to being complete. We just
want to give an example of two resonances which are not considered usually.

Two new difficulties now appear. The first one is that, contrary to
the ${\cal E}_1$-sector case, resonances in the ${\cal E}_2$-sector
are given by zeros of a function $D(\lambda,\mu,.)$ which is no longer
explicit, since we will see that it is the sum of a series. Up to now, 
we can only calculate these zeros approximately, through cutting the series
after the first non trivial term. The zeros of this truncated function
are the approximate values we consider, for the resonances in this
sector.

\noindent
The second difficulty is that this truncated function has several
branch points; as a consequence, it has several analytic
continuations. There is no reason why it should not be the same for
$D(\lambda,\mu,.)$ itself, although it is possible. Therefore, when
considering a zero, one must tell which branch is in question; we give a
notation for the zeros of the various branches later on.

{\bf III.4.2.1 These resonances as zeros of a function
$D(\lambda,\mu,.)$}

In ${\cal E}_2$, looking for the eigenvalues leads to the following
proposition.

{\bf Proposition 3.3} (Eigenvalues of
  $H(\lambda,\mu)\upharpoonright_{{\cal E}_2}$ as zeros of a multivalued
  function) {\it $\lambda$ being fixed and $\mu$ and $z$ being two
parameters satisfying $\mu>0$ and $z\in\bbbr^-$, let ${\cal
D}_{\mu,z}$ be the
Fredholm function (see {\rm Ref. 22, p. 68}) associated with the integral equation
$$
\varphi_{\lambda,\mu}(p)-\lambda^2\int K_{\mu,z}(p,q)\
\varphi_{\lambda,\mu}(q)\  dq=0                \eqno(3.12)
$$
with
$$
K_{\mu,z}(p,q)= {g(p)\overline g(q)\over (z-1-\mu|p|-\lambda^2\
  (T_{\mu,z}g)(p))(z-\mu(|p|+|q|))}       \eqno (3.13)
$$
where
$$
(T_{\mu,z}f)(p)=\int{\overline g(q')f(q')\over z-\mu(|p|+|q'|)} \ dq' \eqno (3.14)
$$
Let $z\rightarrow D(\lambda,\mu,z)$ be the multivalued function
which, for $z\in \bbbr^-$,
equals
${\cal D}_{\mu,z}(\lambda^2)$.

\noindent
In the case where one of the zeros of $D(\lambda,\mu,.)$ is real
negative, this zero is an eigenvalue of $H(\lambda,\mu)$. Let us
denote it by
$\xi_{0,2}(\lambda,\mu)$. The associated eigenvector is in ${\cal
E}_2$, proportional to
$$
\psi_{0,2}^{(0)}(\lambda,\mu):=\ket {1,\varphi_{\lambda,\mu}}+\sqrt
2\ \lambda\ket{0,{g\vee\varphi_{\lambda,\mu}\over
\xi_{0,2}(\lambda,\mu)-\mu(|p_1|+|p_2|)}}     \eqno (3.15)
$$
where $\varphi_{\lambda,\mu}$ is a solution of
$$
\varphi_{\lambda,\mu}(p)-\lambda^2\int K_{\mu,\xi_{0,2}(\lambda,\mu)}(p,q)\
\varphi_{\lambda,\mu}(q)\  dq=0                \eqno(3.16)
$$
} 

\underbar{Proof} $\ket {1,\varphi_{\lambda,\mu}}+\ket{0,
\chi_{\lambda,\mu}}$ is an eigenvector of
$H(\lambda,\mu)\upharpoonright_{{\cal E}_2}$ associated with the eigenvalue
$z$ if and only if
(3.12) holds and $\chi_{\lambda,\mu}=\sqrt 2\ \lambda\big(z-\mu(|p_1|+|p_2|)\big)^{-1}\
g\vee\varphi_{\lambda,\mu}$. According to Fredholm's theory, (3.12) has a
non trivial solution only if $D_{\mu,z}(\lambda^2)=0$. The
Proposition follows from the fact that (3.14) and (3.13) are defined
if $z$ is a negative number. \cqfd
 
\medskip

\noindent
{\it Remark } If $\mu$ is set to $0$ in  (3.12), (3.13) et (3.14), one
finds that (3.12) implies $z=\zeta_{i,2}(\lambda)$. If the continuity
with respect to $\mu$ could be proved, we would get that the limit
 of $\xi_{0,2}(\lambda,\mu)$ when $\mu$ goes to $0$ is $\zeta_{0,2}(\lambda)$.
We do not know how to prove this at the moment, since we do not know all the zeros
of $D(\lambda,\mu,.)$. Nevertheless, the result of the approximate
calculation of Section III.4.2.4 is along this line. That is why we
eventually change notation $\xi_{0,2}(\lambda,\mu)$ for $z_{0,2}^1(\lambda,\mu)$, according to
Notation 2. (The upper index is
explained later on.)

To switch from eigenvalues to resonances, let us take $h\in L^2(\bbbr^2)$, $z$ s.t. $\Im z>0$ and
$\psi:=\ket{1,h}$. Let us introduce $H_{\mu,z}(p,q,\lambda^2)$, the resolvent
kernel of equation (3.12) (see Ref. 22, p. 63). It can be shown that 
$$
(\psi,[z-H(\lambda,\mu)]^{-1}\ \psi)=1-\lambda^2\int H_{\mu,z}(p,q,\lambda^2)\overline h(p)\
h(q)\ dpdq                                                  \eqno (3.17)
$$
We know (Ref. 22,
p. 58 and 63) that the only singular points of $H_{\mu,z}(p,q,.)$
are the
solutions of ${\cal D}_{\mu,z}(\lambda)=0$. The zeros of analytic
continuations of function $D(\lambda,\mu,.)$ of Proposition 3.3 will thus give us 
poles of the left-hand side of (3.17), that is to say resonances.

\noindent
The calculation will be an approximate one. The result for $\mu=1$ is
given by lines two and four of Table 3, in Section III.5.

{\bf III.4.2.2 Zeros of $D(\lambda,\mu,.)$ approached by
zeros of a function
$D^{(1)}(\lambda,\mu,.)$
}

{\bf Proposition 3.4} {\it For $z<0$,
$D(\lambda,\mu,z)=\displaystyle\sum_{n=0}^\infty{(-1)^n\over n!}\ C_n(\lambda,\mu,z)$
with $C_0=1$,
$$
C_n(\lambda,\mu,z)=\lambda^{2n}\int D_n(\mu,z,p_1,\cdots,p_n)\ dp_1\cdots
dp_n                                \eqno (3.18)
$$
and, for $1\leq n\leq3$,
$$
D_n(\mu,z,p_1,\cdots,p_n)=\mu^{-n}\prod_{i=1}^{i=n}{ |g(p_i)|^2\over f(\lambda,\mu,z-\mu
|p_i|)({z\over \mu}-2 |p_i|)}\prod_{i<j}{(|p_i|-|p_j|)^2\over ({z\over
\mu}-(|p_i|+|p_j|))^2}                     \eqno (3.19) 
$$
}

\underbar{Proof} The $D_n$'s are determinants given by Fredholm's
theory. We calculated them for $n\leq 3$; the result is given in (3.19). \cqfd

The analytic structure (poles, branch points) of $D(\lambda,\mu,.)$ is
difficult to determine. Indeed,  $D(\lambda,\mu,.)$ is a series of
terms each of which has a different analytic structure. Moreover, this
structure is not simple. Therefore we will limit ourselves with
replacing the search for zeros of $D(\lambda,\mu,z)$ by the search for
the zeros of the sum of the first two terms of the series:
$$
D^{(1)}(\lambda,\mu,z):=1-C_1(\lambda,\mu,z)      \eqno (3.20)
$$
The first corrections to this approximation are discussed in Appendix
A. From Proposition 3.3, we get, if $\mu\not=0$,
$$
D^{(1)}(\lambda,\mu,z)=1-{2\lambda^2\over \mu}\int_0^\infty
\psi(\lambda,\mu,z,q)\ dq                      \eqno (3.21)
$$
where
$$
\psi(\lambda,\mu,z,q):={g(\mu^{-1}q)^2\over
(z-2|q|)\ f(\lambda,\mu,z-|q|)}              \eqno (3.21')
$$
and $D^{(1)}(\lambda,0,z)={z(z-1)-2\lambda^2\over z(z-1)-\lambda^2}$.

{\bf III.4.2.3 Analytic properties of $D^{(1)}$}

{\bf Proposition 3.5}
{\it For $\mu\not=0$, $D^{(1)}(\lambda,\mu,.)$ has at least three
branch points:
$0$, $z_{0,1}(\lambda,\mu)$ and $z_{1,1}(\lambda,\mu)$.
}

\underbar{Proof} Let us recall that $f(\lambda,\mu,.)$ is analytic in
the complex plane cut along $\bbbr^+$, with a branch point at $0$. The
analytic continuation from the upper half-plane has a pole at
$-i\mu$. An explicit expression of $f$ in the upper half-plane is, for $z\not=i\mu$,
$$
f(\lambda,\mu,z)=z-1+\lambda^2\ {2\mu^3+\mu^2\pi z+2\mu z^2-\pi
z^3+4\mu z^2\big(\log\mu-\log(-z)\big)\over \pi(\mu^2+z^2)^2}    \eqno (3.22)
$$
The continuation clockwise across the cut, which we denote by
$f_+(\lambda,\mu,z)$, is obtained through adding  $4i\pi{\lambda^2\over\mu}
g(\mu^{-1}z)^2$ to the above expression. It is convenient to
rather introduce the function $\hat f(\lambda,\mu,z)$ which coincides
with $f(\lambda,\mu,z)$ in the upper half-plane and has a cut along $i\bbbr^-$.

\noindent
For $\Im z>0$, the poles of $\psi(\lambda,\mu,z,.)$ (see (3.21')) in
the complex plane cut along  $z+i\bbbr^+$
are $\pm i\mu,\ z/2$ and $
q_i(z,\mu,\lambda):=z-z_{i,1}(\lambda,\mu)$. Depending on whether
$\mu$ is smaller or greater than $\mu_c(\lambda)$, the $z$-dependent
poles sit at places schematically shown in Figure 4.
$$
\psboxto(13cm;0cm){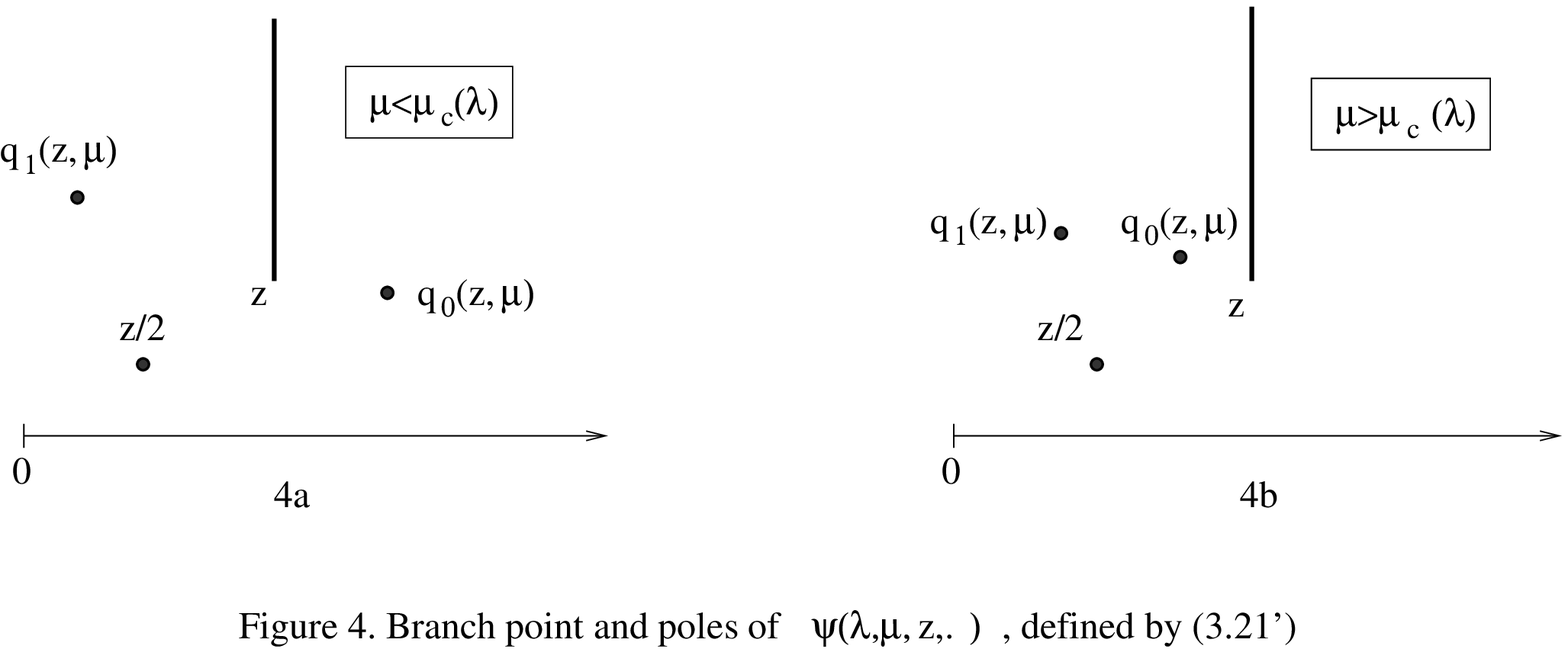}
$$
This follows from the position of $z_{i,1}(\lambda,\mu)$, given by the
curves in Figures 2 and 3.

\noindent
If $\mu=\mu_c(\lambda)$, the pole $q_0(z,\mu)$ coincides with $z$. (In
Figure 4, $z$ has a positive real part, but it could be negative, as
well.)

\noindent
When $z$ enters the lower half-plane along a path $\gamma$ (for
instance the dotted lines in Figures 7 and 8), the
integration path in (3.21) may have to be deformed in order to be kept
away from some of the three poles or from the branch point $z$ of
$\psi(\lambda,\mu,z,.)$. Two different paths $\gamma$ and $\gamma'$ will
not necessarily yield the same result. For instance, if $\gamma$ crosses $z_{i,1}(\lambda,\mu)+\bbbr^+$, $q_i(z,\mu)$ crosses
$\bbbr^+$; the integration path has thus to be deformed, whereas it is
not the case if
$\gamma$ crosses
$z_{i,1}(\lambda,\mu)-\bbbr^+$. As a consequence, $z_{0,1}(\lambda,\mu)$ and
$z_{1,1}(\lambda,\mu)$ are branch points.

\noindent
If $z$ comes to $z_{0,1}(\lambda,\mu)$ (resp. $z_{1,1}(\lambda,\mu)$),
then the pole $q_0(z,\mu)$ (resp. $q_1(z,\mu)$) comes to $0$ and the
integral in (3.21) is singular.

To describe the various branches of $D^{(1)}$ readily, we need a
notation for the homotopy classes of paths in
$X_{\lambda,\mu}:=\bbbc\setminus\{0,z_{0,1}(\lambda,\mu),z_{1,1}(\lambda,\mu)\}$.
(Two paths are homotopic if they can be continuously deformed into one
another, in $X_{\lambda,\mu}$.)
Depending on whether $\mu$ is smaller or greater than
$\mu_c(\lambda)$, the relative position of the three branch points is
different. We refer to Figure 5 for the definition of the fundamental
class of paths $a,a_0,a_1$. The base point $B$ is chosen real and
smaller than $\zeta_{0,2}(\lambda)$.
$$
\psboxto(13cm;0cm){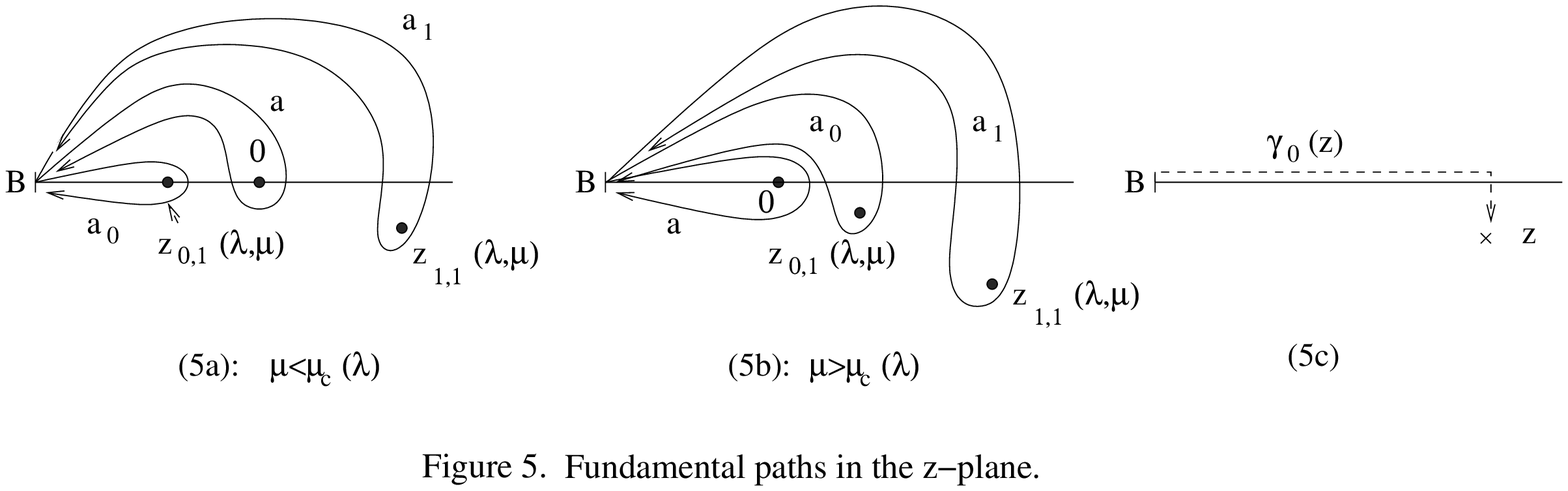}
$$
The path $\gamma_0(z)$ is defined as the polyline going through points
$B$, $B+i\epsilon$, $\Re(z)+i\epsilon$ and ending at $z$, for an
arbitrary $\epsilon>0$. Every path $\gamma$ from $B$ to $z$ can be
expressed by means of $a,a_0,a_1$ and $\gamma_0(z)$. It defines a
homotopy class $[\gamma_0(z)^{-1}\gamma]\in\pi_1(X_{\lambda,\mu})$.
Conversely, with each class $[l]$ of $\pi_1(X_{\lambda,\mu})$, we can
associate the homotopy class of paths $\gamma_0(z)l$ going from $B$ to
$z$.
With each pair consisting of $[l]$ in $\pi_1(X_{\lambda,\mu})$ and $z$,
we can associate the analytic continuation in $z$ of $D^{(1)}(\mu,.)$
along the path $\gamma_0(z)l$. We denote the value at the end of the
path by $D^{(1)}_{[l]}(\mu,z)$. From now on, we will omit to mention
the $\lambda$ variable (which has been fixed to $0.1$.) We denote by 
$z^{[l]}(\mu)$
a point such that $D^{(1)}_{[l]}\big(\mu,z^{[l]}(\mu)\big)=0$. When
$\mu$ varies, we assume that this point varies continuously. It is
denoted by $z_{i,2}^{[l]}(\mu)$ if its limit when $\mu$ goes  to $0$ is 
$\zeta_i$.

These $z_{i,2}^{[l]}(\mu)$ are approximate values for the resonances
we are considering. As regards their physical meaning, we refer to the
short comment in Section III.5.

We now give some values of these functions for various $\mu$ values.

\vfill\eject
{\bf III.4.2.4 Values of resonance $z_{0,2}^{[l]}(\mu)$}

{\it (i) $\mu<6.3662\ 10^{-3}$} 

{\it (a) $[l]=1$.}
For $\mu$ close to $0$, we look for a zero of $D^{(1)}(\mu,.)$ in the
neighbourhood of $\zeta_{0,2}=-1.96\ 10^{-2}$. $D^{(1)}(\mu,.)$ is
well defined by (3.21) in the neighbourhood of every negative real
number. A calculation on a computer yields negative real zeros of this
expression, for $\mu$ small (see Table 2, first column). When $\mu$
increases up to a certain value $\mu_{c,2}^a$, close to
$\mu_c:=\mu_c(10^{-1})$, the same formula still gives a negative real zero.
The first column of Table 2 gives its values for 
 $0\leq\mu\leq 6.3662\ 10^{-3}\simeq\mu_c$. 
$$\vbox{\offinterlineskip\halign{\tv\cc{$#$}\tv&\cc{$\!\!#\!\!$}&\cc{$\!#\!$}&\cc{$\!#\!$}&\cc{$\!\!\!\!#\!$}&\cc{$\!\!\!#\!\!\!$}\tv\cr
\noalign{\hrule}
10^3\mu&10^4z_{0,2}^{{1}}(\mu)&\!\!{C_2\over 2}\big(\mu,z_{0,2}^{1}(\mu)\big)\!\!&{M_3\over
  6}\big(\mu,z_{0,2}^{1}(\mu)\big)&{M_4\over
  24}\big(\mu,z_{0,2}^{1}(\mu)\big)&\partial_zC_1\big(\mu,z_{0,2}^{1}(\mu)\big)\cr
\noalign{\hrule}
0&-196&0&0&0&103.9\cr
0.1&-185.5&1.65\  10^{-3}&3.95\ 10^{-6}&1.04\  10^{-8}&105.3\cr
1&-135.7&13.14\ 10^{-3}&10^{-4}&6.26\ 10^{-7}&116\cr
3&-69.8&0.038&7.93\ 10^{-4}&\ 10^{-5}&143.8\cr
6&-9.3&0.099&7.7\ 10^{-3}&3.68\ 10^{-4}&247.1\cr
6.2&-6.6&0.106&9.3 \ 10^{-3}&5.2\ 10^{-4}&265.3\cr
6.36&-4.4&0.111&1.09\ 10^{-2}&7.1\   10^{-4}&284.2\cr
\! 6.3662&-4.328&0.112&1.10 \ 10^{-2}&7.17\  10^{-4}&285\cr
\noalign{\hrule}}
}$$
\centerline{Table 2. Values of $z_{0,2}^{1}(\mu)$, for some
values of $\mu$ below $\mu_c$.}
\medskip

\noindent
Through comparing Table 2 with Table 1, one sees that this zero is
smaller than $z_{0,1}(\mu)$. In accordance with the notations of the
end of Section III.4.2.3, and
with the upper index $1$ denoting the unity element in $\pi_1(X_\mu)$,
this zero may then be denoted by $z^{{1}}_{0,2}(\mu)$, since the branch
point $z_{0,1}(\mu)$ does not belong to the interval $B
z^{{1}}_{0,2}(\mu)$. Figure 6a shows the three cuts of $D^{(1)}(\mu,.)$ for
$\mu<6.3662\ 10^{-3}$ and this zero $z^{{1}}_{0,2}(\mu)$.

\noindent
It has not been possible to determine the place of $\mu_{c,2}^a$ with
respect to $\mu_c(10^{-1})$. We will come back to this point in
Appendix A when we
 estimate the errors made in the approximation.

{\it (b) $[l]=a_0$: a zero associated with another branch.}
Figure 6a also schematically shows another zero:
$z_{0,2}^{[a_0]}(\mu)$. The notation indicates that this value is now the
zero of the continuation of $D^{(1)}(\mu,.)$ along the path $\gamma':=\gamma_0\big(z_{0,2}^{[a_0]}(\mu)\big)\ a_0$, a path which goes
around the branch point $z_{0,1}(\mu)$. Figure 6b shows the two paths
followed by $q_0(z,\mu)$ when $z$ goes along $\gamma_0\big(z_{0,2}^{(1)}(\mu)\big)$ or
$\gamma'$.
$$
\psboxto(13cm;0cm){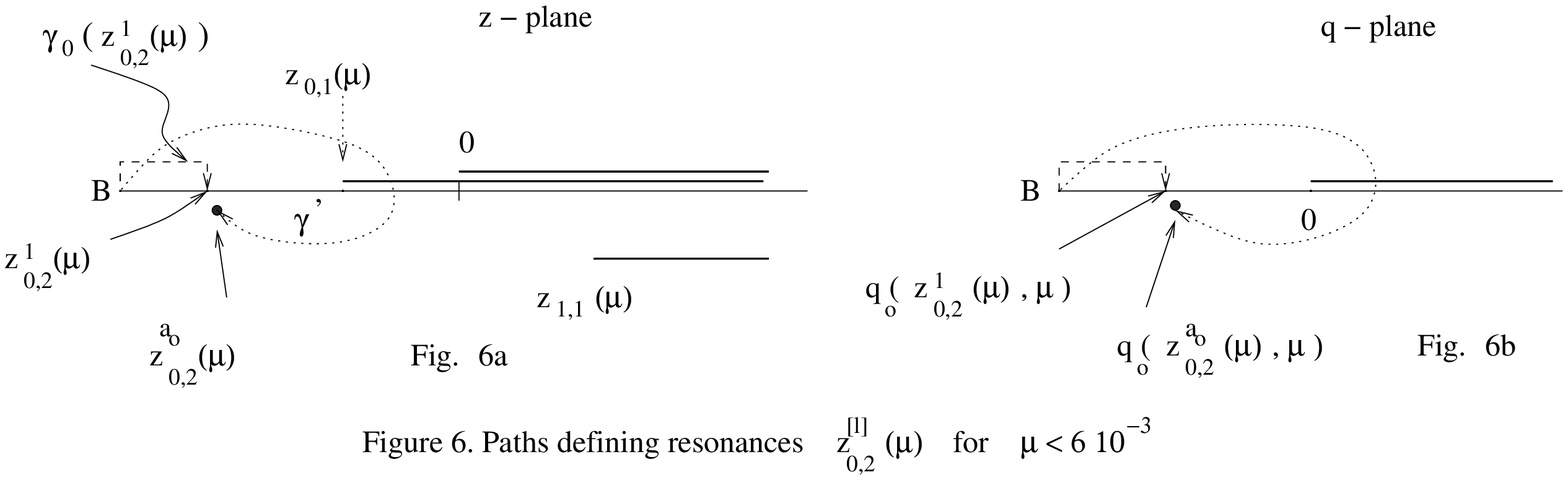}
$$
This $z_{0,2}^{a_0}(\mu)$, also close to
$\zeta_{0,2}$ for $\mu\simeq0$, is a zero of
$D^{(1)}_{a_0}(\mu,.)$ which, for $\Re(z)<0$ (and also for
$\Re(z)>0,\Im(z)>0$), reads

$$\displaylines{
D^{(1)}_{a_0}(\mu,z)=\hfill\cr
\hfill 1-{2\lambda^2\over \mu}\int_0^\infty {g(\mu^{-1}q)^2\over
(z-2|q|)\ f(\mu,z-|q|)}
\ dq  +4i\pi{\lambda^2\over\mu} {g\big(q_0(z,\mu)/\mu\big)^2\over
\big(z-2 q_0(z,\mu)\big)\partial_z f(z_{0,1}(\mu,\lambda),\mu,\lambda)
}\hfill (3.23)      }$$
with $\lambda=0.1$. This zero is no longer real.
\smallskip

{\it (ii)} For $\mu\simeq\mu_c$, things are not clear.

{\it (iii)} For $\mu> 7\ 10^{-3}$, the branch point $z_{0,1}$
of $D^{(1)}$ is in
$\{z;\Re (z)>0,\Im (z)<0\}$ (see Figure 2); it is shown in Figure
7. Among the various analytic continuations of $D^{(1)}$ across
$\bbbr^+$, we consider $D^{(1)}_1$. Expressions
are given below (formula (3.24) and the two last lines of the
section). The zero of $D^{(1)}_1$, denoted by $z_{0,2}^1(\mu)$,
follows the curve of Figure 7 when $\mu$ varies from $7\  10^{-3}$ to
$1$. For $\mu=1$, the way $D^{(1)}$ is analytically continued to the
zero of $D_1^{(1)}$ is shown by the dotted line.

\vskip 0.3cm
$$
\psannotate{  \psboxto(0cm;4cm){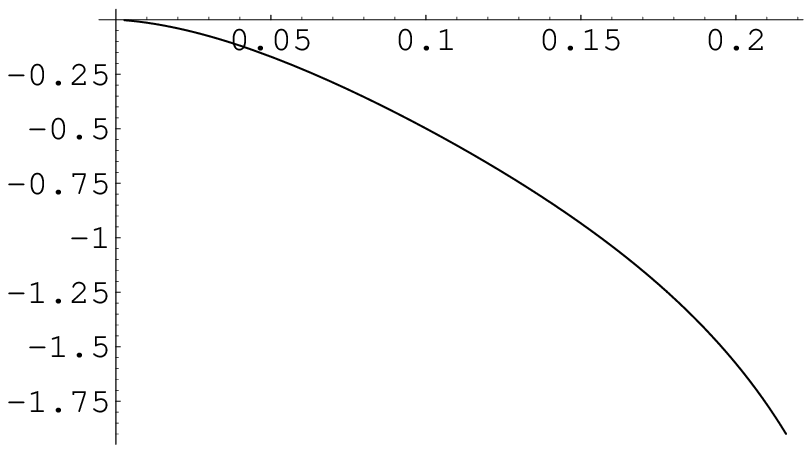}  }  {\at{3.2cm}{1.6cm}{$z_{0,1}(0.1,1)$}    \at{4.1cm}{1.9cm}{$\bullet$}  \at{1.5cm}{3.9cm}{$_\swarrow\mu=0.007$}
\at{5.1cm}{0.2cm}{$\mu=1\rightarrow$}  \at{-1cm}{-0.5cm}{Figure
7. $z_{0,2}^1(\mu)$, in the complex plane, for $\mu\in[0.007,1]$
}\at{4.5cm}{4.3cm}{$\rightarrow$}\at{-1.1cm}{4.2cm}{
$.\!\ .\!\  .\!\  .\!\  .\!\ .\!\  .\!\  .\!\  .\!\ .\!\  .\!\  .\!\  .\!\ .\!\  .\!\  .\!\  .\!\ .\!\  .\!\  .\!\  .\!\ .\!\  .\!\  .\!\  .\!\ .\!\  .\!\  .\!\  .\!\ .\!\  .\!\  .\!\  .\!\ .\!\  .\!\  .\!\  .\!\ .\!\  .\!\  .\!\  .\!\ .\!\  .\!\  .\!\  .\!\ .\!\  .       $ }\at{6.4cm}{4.1cm}{:}\at{6.4cm}{3.85cm}{:}\at{6.4cm}{3.60cm}{:}\at{6.4cm}{3.35cm}{:}\at{6.4cm}{3.1cm}{:}\at{6.4cm}{2.85cm}{:}\at{6.4cm}{2.6cm}{:}\at{6.4cm}{2.35cm}{:}\at{6.4cm}{2.1cm}{:}\at{6.4cm}{1.85cm}{:}\at{6.4cm}{1.60cm}{:}\at{6.5cm}{1.60cm}{$\downarrow$}\at{6.4cm}{1.35cm}{:}\at{6.4cm}{1.1cm}{:}\at{6.4cm}{0.85cm}{:}\at{6.4cm}{0.60cm}{:}\at{6.4cm}{0.35cm}{:} }
$$
\vskip 0.5 cm

\noindent
For $\mu=7\  10^{-3}$, $z_{0,2}^{1}(\mu)=2.8\
10^{-4}-2.4\ 10^{-5}\ i$, close to $0$. Values of $z_{0,2}^{1}(\mu)$ for $\mu$ between $6.37\
10^{-3}$ and $7\  10^{-3}$ are difficult to get. Actually, their
determination is not useful since $z_{0,2}^{1}(\mu)$ is only an
approximate value of the exact resonance. We have
$z_{0,2}^{1}(0.1,1)=0.216-1.9\ i$. For $\Im z<0$,
$\Re (z)>0$ and $\mu>7.35\ 10^{-3}$, still with $\lambda=0.1$,
$$\displaylines{
D^{(1)}_1(\mu,z)=1-{2\lambda^2\over\mu}\Big(\int_{[0,\Re (z)]\cup [\Re
(z),z]}{g^2(\mu^{-1}q)\over (z-2q)\ f_+(\mu,z-q)}+\hfill\cr\hfill \int_{[0,\Re (z)]\cup [\Re
(z),z]}{g^2(\mu^{-1}q)\over (z-2q)\ f(\mu,z-q)}+i\pi
g^2({z\over 2\mu})\ f_+(\mu,z/2)^{-1}+\hfill\cr\hfill 2i\pi
g^2\big({z-z_{0,1}(\mu)\over\mu}\big)\  (2 z_{0,1}(\mu)-z)^{-1}\ 
\partial_z f_+(\mu,z_{0,1}(\mu))^{-1}     \Big)  \hfill(3.24)
}$$
This formula is derived from (3.21) in the following way: when $z$
enters the quadrant $\{z;\Re (z)>0,\Im (z)<0\}$, the cut in Figure 4b
drags the integration contour along, which yields the first two
terms. The last two terms come from the residues of the poles $z/2$
and $q_0(z,\mu)$ which cross $\bbbr^+$ ($q_1(z,\mu)$
does not cross $\bbbr^+$). For $10^3\mu\in[7,7.3]$, the expression of
$D^{(1)}_1(\mu,z)$ does not contain the residue term at $q_0$ since
$\gamma_0\big(z_{0,2}^1(\mu)\big)$ does not cross $z_{0,1}(\mu)+\bbbr^+$.

{\bf III.4.2.5 Values of resonance $z_{1,2}^1(\mu)$}

It can be shown numerically that the zero of $D_1^{(1)}(\mu,.)$ which
tends to $\zeta_{1,2}$ when $\mu$ tends to $0$, denoted by
$z_{1,2}^1(\mu)$, is a zero in $\{z;\Re(z)>0,\Im(z)<0\}$ of
$\Delta(\mu,.)$, where
$$\displaylines{
\Delta(\mu,z):=1-{2\lambda^2\over\mu}\Big(\int_{[0,\Re (z)]\cup [\Re
(z),z]}{g^2(\mu^{-1}q)\over (z-2q)\ f_+(\mu,z-q)}+\hfill\cr\hfill \int_{[0,\Re (z)]\cup [\Re
(z),z]}{g^2(\mu^{-1}q)\over (z-2q)\ f(\mu,z-q)}+i\pi
g^2({z\over 2\mu})\ f_+(\mu,z/2)^{-1}+\hfill\cr\hfill 2i\pi
g^2\big({z-z_{0,1}(\mu)\over\mu}\big)\  (2 z_{0,1}(\mu)-z)^{-1}\ 
\partial_z f_+(\mu,z_{0,1}(\mu))^{-1}+ \hfill\cr\hfill 2i\pi
g^2\big({z-z_{1,1}(\mu)\over\mu}\big)\  (2 z_{1,1}(\mu)-z)^{-1}\ 
\partial_z f_+(\mu,z_{1,1}(\mu))^{-1}    \Big)  \hfill              (3.25)
}$$
$\Delta(\mu,z)$ is the expression of $D_1^{(1)}(\mu,z)$ in the
neighbourhood of the considered zero, but not everywhere in the lower
half-plane; for example, these two functions differ at a point $z$
such that $\Im(z)<\Im\big(z_{0,1}(\mu)\big)$. The variation 
of $z_{1,2}^1(\mu)$ with $\mu$ is given by Figure 8.
\vskip 1cm
$$
\psannotate{  \psboxto(0cm;4cm){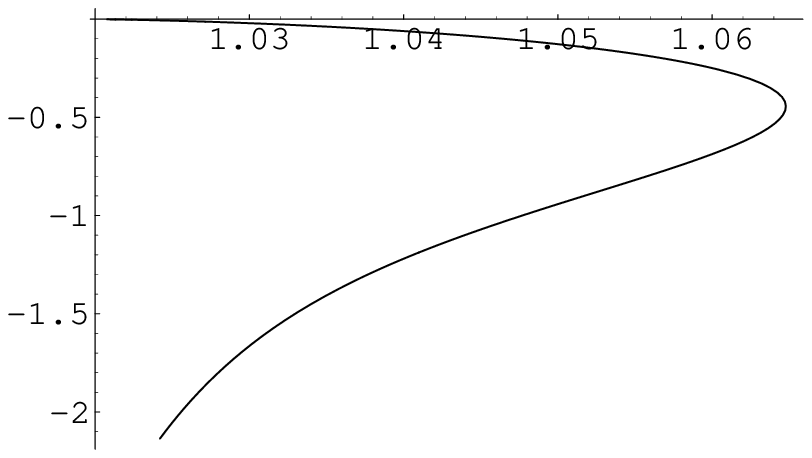}  }
{ \at{3cm}{4.3cm}{$\rightarrow$}\at{-1.1cm}{4.2cm}{
$.\!\ .\!\  .\!\  .\!\  .\!\ .\!\  .\!\  .\!\  .\!\ .\!\  .\!\  .\!\  .\!\ .\!\  .\!\  .\!\  .\!\ .\!\  .\!\  .\!\  .\!\ .\!\  .\!\  .\!\  .\!\ .\!\  .\!\  .\!\  .\!\ .\!\  .\!\  .\!\  .\!\ .\!\  .     $ }\at{4.4cm}{4.1cm}{:}\at{4.4cm}{3.85cm}{:}\at{4.4cm}{3.60cm}{:}\at{4.4cm}{3.35cm}{:}\at{4.4cm}{3.1cm}{:}\at{4.4cm}{2.85cm}{:}\at{4.5cm}{2.6cm}{$\downarrow$}\at{4.4cm}{2.6cm}{:}\at{4.4cm}{2.35cm}{:}\at{4.4cm}{2.1cm}{:}  \at{4.4cm}{1.9cm}{$\leftarrow \mu=1$}  \at{1cm}{3.9cm}{$_\downarrow\mu=0$}
\at{1.55cm}{0.25cm}{$\leftarrow\mu=2$}  \at{0cm}{-0.4cm}{Figure
8. $z_{1,2}^1(\mu)$, in the complex plane, for $\mu\in
[3\  10^{-4},2]$ }  }
$$
\vskip1cm
 This resonance starts from $2^{-1}(1+\sqrt{1+8 \lambda^2})=1.01962$
for $\mu=0$ and goes through $1.043-1.127\ i$ for $\mu=1$. We note
that this value is much farther from the real axis than the resonance
$z_{1,1}(0.1,1)$ of sector ${\cal E}_1$. As in Figure 7, the dotted
line in Figure 8 indicates the path of the analytic continuation of
$D^{(1)}$ to the zero (of  $D^{(1)}_1$), for $\mu=1$. The origin is a
branch point. The other two branch points, $z_{0,1}(0.1,1)$ and
$z_{1,1}(0.1,1)$, whose values are recalled in Table 3, cannot
be drawn at the scale on the figure.

\vfill\eject
{\bf III.5 Conclusion of Section III}

Table 3 gathers the four results we obtained for $\lambda=0.1$ and
$\mu=1$. It gives four resonances among all those of the physical (i.e
$\mu=1$) Hamiltonian, when the function $g$ is given by (3.8).
$$\vbox{\openup 1mm\offinterlineskip\halign{\tv\cc{$#$}\tv&\cc{$#$}\tv\cr
\noalign{\hrule}
z_{0,1}&0.13-1.97\ i\cr
z_{0,2}^1&0.216-1.9\ i\cr
z_{1,1}&0.997-0.010\ i\cr
z_{1,2}^1&1.043-1.127\ i\cr
\noalign{\hrule}}
}$$
\centerline{Table 3. Four resonances (approximate values for two of them),}
\vskip-0.1cm
\centerline{ for $\lambda=0.1$ and $\mu=1$}

\smallskip

\noindent
We see that the usual resonance $z_{1,1}$ is the only one near
the real axis. However, for small values of $\mu$, this is no longer
true, as it can be seen from Figures 2, 7 and 8, or Table
2. Therefore, resonances for small $\mu$ may play an important
role. Now, in the harmonic-oscillator example mentioned in the
introduction, what plays the role of parameter $\mu$ is $\delta^{-1}$,
the inverse of the spatial extension of the states. Thus, for extended
states, $\mu$ is small, and we recover our motivation for 
the study of all the resonances of atom-field Hamiltonians.

One should of course discuss which of these resonances have a physical
meaning and make this meaning precise. One should also find actual
situations in which these resonances or eigenvalues can be seen
readily. Unfortunately, some more work has to be done. We think that
 the physical meaning
should appear quite easily in the case of eigenvectors of the Hamiltonian. In ${\cal E}_1$ (see
(3.10), Table 1 and Figure 2), we have an example of such a state. As
regards the restriction of $H(\lambda,\mu)$ to ${\cal E}_2$, we did
found approximate real values for the resonances. But we have not shown
that (real) eigenvalues do exist. The study has to be carried
on. Now, as pure resonances are concerned, we think that their study
cannot simply be an
academic question,
since eigenvectors change continuously into resonances when parameters
of the physical system are varied.

In any case, the 2-level model is perhaps too
simple to find a concrete application. 
The next section is a step towards more realistic models. 

\medskip
{\bf IV Hybrid states for a three-level atom coupled to photons}

To be in a position to describe mixed states in more realistic models, we
not only must be able to consider several photons, as in Section III, but
we must also be able to consider several atomic or molecular levels. 
The present
section is a preliminary study devoted to mixed states when $S$ is a
three-level atom coupled to the radiation by a Hamiltonian of type
(2.2).

The reason why the study is only a preliminary one is that there come
some additional difficulty, together with those already mentioned in
Section III: for $\mu\geq 0$, none of the spaces with a bounded
excitation number is stable by the evolution operator, contrary to
the two-level case. This is due to the possibility of
the transition $\ket 0+photon\rightarrow\ket 2$ ($\ket 2$ is the
second excited state), through which the
excitation number may increase. For this reason, the determination of
the Hamiltonian eigenvalues is already not simple for
$\mu=0$. Since we assume that the structure of the resonance set will
roughly be conserved when $\mu$ takes non zero values, we must
study the ($\mu=0$)-problem first. This is the subject of this Section
IV.\break The displacement of these resonances when $\mu$ becomes non-zero
will not be examined in the paper.

\noindent
A typical result is illustrated by Figure 12.

\medskip
{\bf IV.1 Notations and Hamiltonian}

The atom has three levels with energies $e_0=0,\ e_1$ and $e_2$,
corresponding to states $\ket 0,\ket 1$ and $\ket
2$. $f_{01},f_{12},f_{02}$ being three normed functions in
$L^2(\bbbr)$, the Hamiltonian is
$$
H(\lambda,\mu):=H_0(\lambda,\mu)+\lambda_{02}\ V                 \eqno(4.1)
$$
with
$$\displaylines{
H_0(\lambda,\mu):=\big(e_1\dyad 1 1 +e_2\dyad 2 2\big)\otimes 1 +\mu\ 1\otimes
H_{\rm rad}+\hfill\cr\hfill\lambda_{01}\Big(\dyad 1 0\otimes\  c(\bar
f_{01})+\dyad 0 1\otimes\  c^*(f_{01})\Big)+\lambda_{12}\Big(\dyad 2
1\otimes\ 
c(\bar f_{12})+\dyad 12\otimes\  c^*(f_{12})\Big)        \hfill(4.2)
}$$
$$
V:=\dyad 2 0\otimes\ 
c(\bar f_{02})+\dyad 0 2\otimes\  c^*(f_{02})          \eqno(4.3)
$$
$H_{\rm rad}$ is as in Section III.1. Transition  $\ket
0\rightarrow\ket2$ is distinguished from $\ket 0\rightarrow\ket1$ and
$\ket 1\rightarrow\ket2$ for a physical reason and also to prepare a
perturbative calculus.

We use the following notations, modeled on those of Section III.1. As
regards the photons, we denote the Fock space by ${\cal F}$, the
$n$-photon 
space by ${\cal F}_n$, the vacuum state by $\Omega$, and we set

$ {\cal H}_{\rm rad}^{(2)}$: the space spanned by
$f_{01},f_{12}$,
$ {\cal F}^{(2)}_n:= \big({\cal H}_{\rm rad}^{(2)}\big)^{\vee^n}$: the
$n$
photon-state space, but \indent each photon being restricted to be in $ {\cal H}_{\rm rad}^{(2)}$ and
$ {\cal F}^{(2)}:=\build\oplus_n^{}  {\cal F}^{(2)}_n$.

$ {\cal H}_{\rm rad}^{(3)}$: the space spanned by $f_{01},f_{12},f_{02}$,
$ {\cal F}^{(3)}_n:= \big({\cal H}_{\rm rad}^{(3)}\big)^{\vee^n}$ (
each photon is restricted \indent to be in $ {\cal H}_{\rm
rad}^{(3)}$) and
$ {\cal F}^{(3)}:=\build\oplus_n^{} {\cal F}^{(3)}_n$.

\noindent
As regards the atom-photon system, ${\cal E}_n$ and ${\cal E}$ have
been defined in Section III.1. We set

${\cal E}_1^{(k)}:=\ket 0\otimes {\cal F}_1^{(k)}  \oplus \ket 1\otimes
{\cal F}_{0}^{(k)} ,\quad k=2,3 $ 

${\cal E}_n^{(k)}:=\ket 0\otimes {\cal F}_n^{(k)}  \oplus \ket 1\otimes
{\cal F}_{n-1}^{(k)}       \oplus \ket 2\otimes {\cal F}_{n-2}^{(k)},\
\quad k=2,3,\quad n\geq 2$   

${\cal E}^{(k)}:=\build\oplus_{n=0}^{\infty} {\cal E}_n^{(k)}$

$\lambda:=(\lambda_{01},\lambda_{12},\lambda_{02})$

\noindent
We also introduce $s_0:=(f_{01},f_{02}),\ s_1:=(f_{01},f_{12}),\
s_2:=(f_{02},f_{12})$.

\noindent
We will use the letter $\phi$ to indicate eigenvectors of
$H_0(\lambda,0)$; a priori, they depend on $\lambda_{01}$ and
$\lambda_{12}$. We will use $\chi$ to indicate eigenvectors of $H(\lambda,0)$;
they also depend on $\lambda_{02}$.

We aim at getting the eigenvalues of $H(\lambda,0)$, an operator which
we simply write  $H(\lambda)$, from now on. When the variable $\mu$ is
not mentioned, it will be assumed to be $0$.

\noindent
To this end, we take up the idea mentioned in Section II consisting in
perturbing $H(0)$ through introducing the interaction step by
step. The doubly infinite degeneracy of the eigenvalues $e_0$,
$e_1$ and $e_2$ of $H(0)$, due to the arbitrariness of the number of
photons and the arbitrariness of
the state of each photon in the corresponding eigenvectors, is
partially removed at each step. The first perturbation will be the
addition of the $\lambda_{01}$ 
and $\lambda_{12}$ terms of (4.2) to $H(0)$.
It is described in  Section IV.2. The second
perturbation will be the supplementary addition of $\lambda_{02}\ V$.
It is described in Section
IV.3.

\vfill\eject
{\bf IV.2 Perturbation with respect to $\lambda_{01}$ and
$\lambda_{12}$. First splitting of $e_0$, $e_1$ and $e_2$}

Here we are interested in $H_0(\lambda):=H_0(\lambda,0)$; the
interaction $\lambda_{02}V$ is switched off. First, in ${\cal E}_0$,
$\ket{0,\Omega}$, is an eigenvector associated with the eigenvalue $0$.

\smallskip
{\bf IV.2.1 Three eigenvectors of $H_0(\lambda)$ in the 1-excitation
space  ${\cal E}_{1}^{(2)}$ and the three associated eigenvalues}

The space ${\cal E}_1^{(2)}$, three-dimensional, is invariant by
$H_0(\lambda)$. It is the direct sum of the eigen-subspaces $\ket 1\otimes {\cal
  F}^{(2)}_0$ and $\ket 0\otimes {\cal F}^{(2)}_1$ of $H_0(0)$,
associated with the eigenvalues $e_1$ and $e_0$ respectively, and ${\rm dim.}\  {\cal F}^{(2)}_0=1$ and ${\rm dim.}\  {\cal
  F}^{(2)}_1=2$. The first perturbation will shift $e_1$ and split
$e_0$ into two eigenvalues, as it is represented in the first two
columns of Figure 9.
\setbox1=\psboxto(5.8cm;0cm){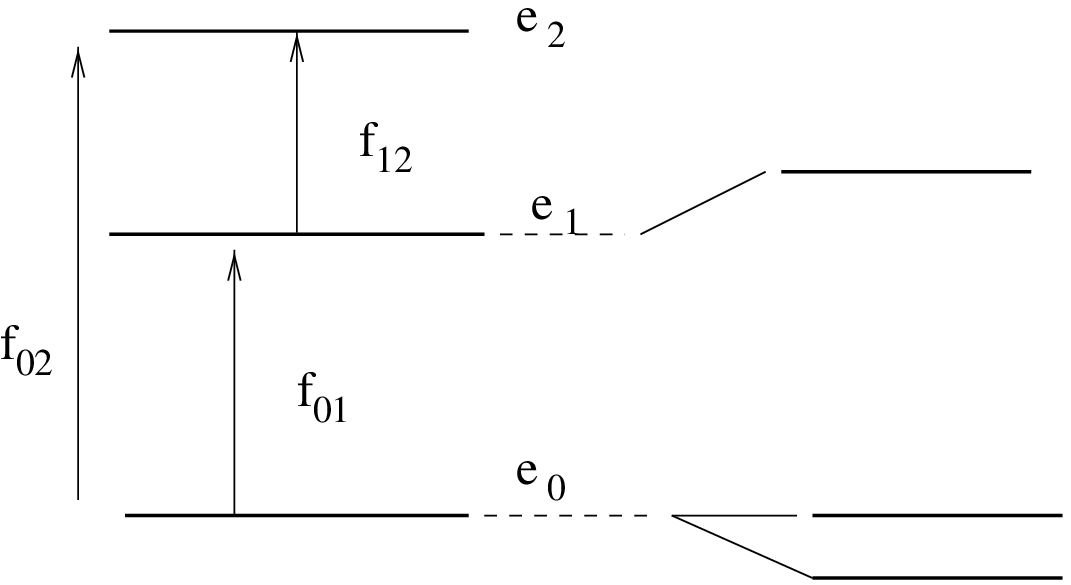}
\setbox2=\vbox{\hbox{\eightrm levels of the atom}\hbox{\eightrm
decoupled from the field}}
\setbox3=\vbox{\hbox{\eightrm 1-excitation}\hbox{\eightrm mixed levels}}
\setbox4=\vbox{\hbox{\quad\quad 1}\vskip 1.6cm\hbox{\quad\quad
    2}\vskip 0.65cm\hbox{\eightrm dimension of the field}\hbox{\eightrm
    state-space (apart}\hbox{\eightrm from degeneracy)}}
\setbox5=\vbox{\hbox{$\zeta_{1,1}$}\vskip1.25cm\hbox{$(\zeta_{0,1})_2$}\hbox{$(\zeta_{0,1})_1$}\vskip1cm\hbox{\eightrm
    eigen-}\hbox{\eightrm values}}
\setbox6=\vbox{
  \hbox
  {\hbox{$\phi_{1,1}^{(0)}$}\quad\raise 1mm\hbox{$\ \ _{\ket{1,0},\ \ket{0,1}}$}
     }\vskip1cm\hbox{$(\phi_{0,1}^{(0)})_2$\quad\ \ 
  $_{\ket{0,1}}$}\hbox{ $(\phi_{0,1}^{(0)})_1$
 {\ $_{\ket{1,0},\ \ket{0,1}}$}  }\vskip 1cm      \hbox{\eightrm
 eigenvectors and} \hbox{\eightrm
  their components$^{(1)}$}}


\setbox7=\hbox{\box2\hskip 1.33cm\box3}

\setbox8=\vbox{\box1\vskip 1.2cm\box7}

\setbox9=\line{\box8\quad\box5\quad\box4\quad\box6}

\vbox{\box9\medskip\hbox {\centerline{Figure 9. Levels associated with
    eigenvectors of $H_0(\lambda)$ in ${\cal
    E}_1^{(2)}$,}} \hskip 3cm\hbox{ therefore without
    any spectator-photon}}

\footnote{}{$^{(1)}$ \eightrm $\ket{i,n}$ schematically indicates a
$n$-photon state with the atom in state $i$ .}
\vskip 1cm
\noindent

The third column gives the notation of the perturbed eigenvalues. In
the fourth column, we recalled the above mentioned number $1$ and $2$,
which are the dimensions of the projections of ${\cal
  E}_1^{(2)}$ on $\ket i\otimes {\cal F}^{(2)}$. They also are the sum
of the dimensions of the eigenspaces of $H(\lambda)\upharpoonright_
{{\cal E}_1^{(2)}}$ associated with eigenvalues which tend to $e_i$
when $\lambda$ tends to $0$. ``Apart from degeneracy'' indicates that in
considering ${\cal E}_1^{(2)}$ we do not take account of the additional
degeneracy of the eigenvalues due to photons in states orthogonal to
$f_{01}$ and $f_{12}$; we call these photons spectator-photons. This
degeneracy is explained in Proposition 4.2 and is removed by the
second perturbation studied in Section IV.3.

\noindent
The eigenvalues obtained through the first perturbation and the
associated eigenvectors are given by the following proposition.

{\bf Proposition 4.1} {\it In ${\cal E}_1^{(2)}$, if
$f_{01}\not=f_{12}$, $H_0(\lambda)$ has three eigenvalues:

$\big(\zeta_{0,1}\big)_1(\lambda_{01},\lambda_{12})=2^{-1}(e_1-\sqrt{e_1^2+4\lambda_{01}^2})$
\hfill{\rm (4.4)}

$\big(\zeta_{0,1}\big)_2(\lambda_{01},\lambda_{12})=0$    \hfill{\rm (4.5)}

$\zeta_{1,1}(\lambda_{01},\lambda_{12})=2^{-1}(e_1+\sqrt{e_1^2+4\lambda_{01}^2})$
                       \hfill{\rm (4.6)}
\smallskip
The associated normed eigenvectors are
\smallskip
$\big(\phi_{0,1}^{(0)}\big)_1(\lambda_{01},\lambda_{12})=\big(1+(\zeta_{0,1})_1^{-2}\lambda_{01}^2\big)^{-1/2}\big(\ket{1,\Omega}+(\zeta_{0,1})_1^{-1}\lambda_{01}\ket{0,f_{01}}\big)$  \hfill{\rm (4.7)}

$\big(\phi_{0,1}^{(0)}\big)_2(\lambda_{01},\lambda_{12})=(1-|s_1|^2)^{-1/2}\ket{0,f_{12}-s_1f_{01}}$.  \hfill{\rm (4.8)}

$\phi_{1,1}^{(0)}(\lambda_{01},\lambda_{12})=\big(1+(\zeta_{1,1})^{-2}\lambda_{01}^2\big)^{-1/2}\big(\ket{1,\Omega}+(\zeta_{1,1})^{-1}\lambda_{01}\ket{0,f_{01}}\big)$  \hfill{\rm (4.9)}}
\smallskip
The calculus is straightforward since the dimension of ${\cal
E}_1^{(2)}$ is three. \cqfd

To the second order with respect to $\lambda$, we get $(\zeta_{0,1})_1=-{1\over
  e_1}\lambda_{01}^2$ and $\zeta_{1,1}=e_1+{1\over
  e_1}\lambda_{01}^2$. Note that the perturbed eigenvalues (4.4) to
  (4.6) do not depend on $\lambda_{12}$. They tend to $0$, $0$ and
  $e_1$, respectively, if $\lambda_{01}$ tends to $0$. The notation
  $\zeta_{i,p}$ is thus in accordance with the rules stated in Section
  II, as far as the first index is concerned; index $i$ refers to the
  unperturbed level in the following way: $
\lim_{\lambda_{01}\rightarrow 0}(\zeta_{0,1})_j=0\ ,\quad
\lim_{\lambda_{01}\rightarrow 0}\zeta_{1,1}=e_1
$.
To distinguish perturbed eigenvalues which tend to the same $e_i$ if
$(\lambda_{01},\lambda_{12})\rightarrow(0,0)$, we chose to put the
  index $p$, indicating that the eigenvectors are in ${\cal E}_p$ (as
  in Section III.2). Since several eigenvalues with the same indices $i$
  and $p$ may still have the same limit when
  $(\lambda_{01},\lambda_{12})$ tends to $(0,0)$, an additional index
  $j$ is used to number them.

We have $\big(\phi_{0,1}^{(0)}\big)_1(\lambda_{01},\lambda_{12})\rightarrow -\ket{0,f_{01}}$ and
$\phi_{1,1}^{(0)}(\lambda_{01},\lambda_{12})\rightarrow\ket{1,\Omega}$,
if $\lambda_{01}$ tends to
$0$.

If $f_{01}=f_{12}$, the dimension of ${\cal E}_1^{(2)}$ is two and the
eigenvector $\big(\phi_{0,1}^{(0)}\big)_2$ disappears.

As we said before, these eigenvalues are actually infinitely
degenerated; indeed, adding photons orthogonal to $\{f_{01},f_{12}\}$
gives an eigenstate with the same energy, since the energy of the
photons is not taken into account in $H_0(\lambda)$. Let us state this
fact precisely, with some notations which will be useful later on.

\smallskip
{\bf IV.2.2 Other eigenvectors of $H_0(\lambda)$ in the
$(n+1)$-excitation space ${\cal E}_{n+1}$,
$n>0$, associated with the same eigenvalues}.

{\bf Proposition 4.2} {\it Let $g_1,g_2,\cdots$ be an
  orthonormal basis of functions
  orthogonal to $f_{01}$ and $f_{12}$. 

(i) Let
${\cal G}_{0,1,1}$ be the subspace of ${\cal E}$ spanned
  by the normed vectors 
$$
(\phi_{0,n+1}^{(n_1,n_2,\cdots)})_1:=\prod_{i=1}^\infty\ (n_i!)^{-{1\over
    2}}\big(1\otimes c^*(g_i)\big)^{n_i}\
(\phi_{0,1}^{(0)})_1          \eqno (4.10)
$$

\noindent
where the $n_i$'s are $k$ non-negative integers and $k$ is arbitrary. These
vectors are in ${\cal E}_{n+1}$\break if $n=\displaystyle\sum_{i=1}^k
n_i$. For generic values of $\lambda_{01}$ and $\lambda_{12}$,
${\cal G}_{0,1,1}$ is the eigenspace of $H_0(\lambda)$ associated with
 eigenvalue $ (\zeta_{0,1})_1$.

(ii) If $f_{01}\not=f_{12}$, let us set $g_0:=(1-|s_1|^2)^{-{1\over 2}}\
(f_{12}-s_1f_{01})$ and let $f_{01}^\bot$ be the subspace of ${\cal H}_{\rm
rad}$ orthogonal to $f_{01}$
, spanned by the $g_i$'s, $i=0,1,\cdots$. The eigenspace of
$H_0(\lambda)$ associated with eigenvalue
$(\zeta_{0,1})_2$ is $\ket{0}\otimes{\cal F}(f_{01}^\bot)$, where ${\cal
F}(f_{01}^\bot)$ is the Fock space built with $f_{01}^\bot$. We set
$$
(\phi_{0,n+1}^{(n_0,n_1,\cdots)})_2:=\prod_{i=0}^\infty\ (n_i!)^{-{1\over
    2}}\big(1\otimes c^*(g_i)\big)^{n_i}\
(\phi_{0,1}^{(0)})_2          \eqno (4.11)
$$

(iii) Lastly, let
${\cal G}_{1,1}$ be the subspace of ${\cal E}$ spanned
  by the normed vectors
$$
\phi_{1,n+1}^{(n_1,n_2,\cdots)}:=\prod_{i=1}^\infty\ (n_i!)^{-{1\over
    2}}\big(1\otimes c^*(g_i)\big)^{n_i}\
\phi_{1,1}^{(0)}           \eqno (4.12)
$$
For generic values of $\lambda_{01}$ and $\lambda_{12}$, ${\cal
  G}_{1,1}$ is the eigenspace of $H_0(\lambda)$ associated with
eigenvalue
$\zeta_{1,1}$. }

That $\prod_{i=1}^\infty\big(1\otimes c^*(g_i)\big)^{n_i}\
(\phi_{0,1}^{(0)})_1$, $\prod_{i=0}^\infty\big(1\otimes c^*(g_i)\big)^{n_i}\
(\phi_{0,1}^{(0)})_2$ and $\prod_{i=1}^\infty\big(1\otimes c^*(g_i)\big)^{n_i}\
\phi_{1,1}^{(0)}$ are eigenvectors  follows from the fact that $[1\otimes c^*(g),H_0(\lambda)]=0$ if $g$ is
orthogonal to ${\cal H}_{\rm rad}^{(2)}$. In Appendix B1, we explain
how the whole eigenspace can be determined, for each of the three
eigenvalues. \cqfd

Let $g$ (which is no longer function (3.8)) be a linear combination of
the $g_i$'s. Proposition 4.2 may be stated in the following terms: the
three eigenvalues are twice infinitely degenerated; firstly through
the number of spectator-photons (the $n$ variable), and secondly
through the infinity of possible states for each
spectator-photon ( the $g$ variable). In the level diagrams, if we
symbolise the degeneracy of an eigenvalue (due to the possibility of one
spectator-photon in the eigenvector) by a dotted line, then in the
case where at most one spectator-photon is present, the degeneracy of
the levels may be represented in the following way: 
\setbox21=\psboxto(5.8cm;0cm){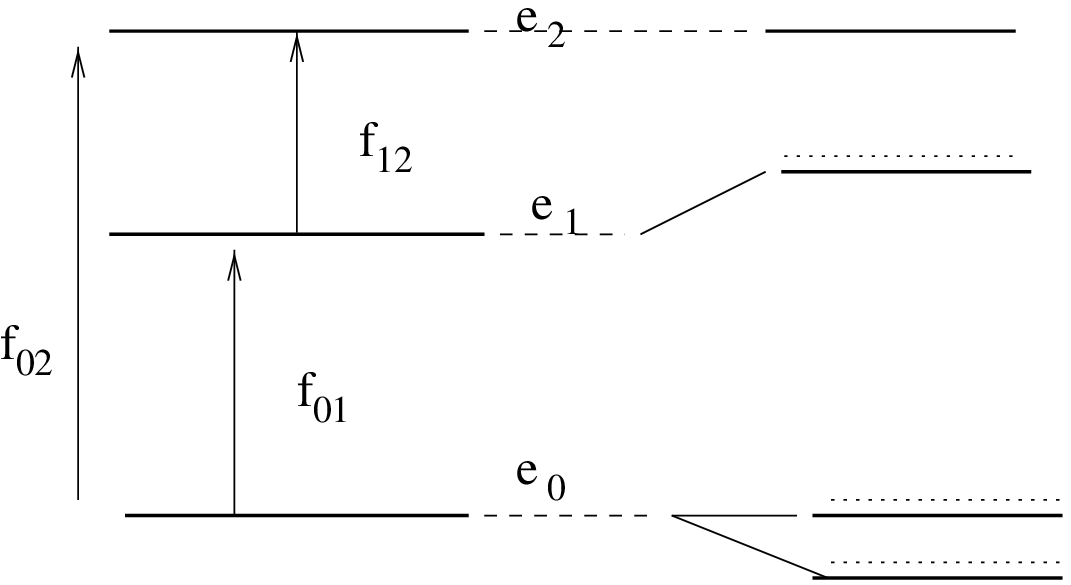}

\setbox22=\vbox{\hbox{\eightrm levels of the decoupled}\hbox{\eightrm
    atom}}

\setbox23=\vbox{\hbox{\eightrm two-excitation}\hbox{\eightrm mixed levels,}\hbox{\eightrm with spectator-}\hbox{\eightrm
    photons{$^{(1)}$}}}

\setbox00=\vbox{\hbox{$\zeta_{1,1}$}\vskip1.4cm\hbox{$(\zeta_{0,1})_2$}\hbox{$(\zeta_{0,1})_1$}\vskip1.53cm\hbox{\eightrm
    eigen}\hbox{\eightrm values}}

\setbox24=\vbox{\hbox{\eightrm\quad\quad \ 1}\vskip 0.2cm\hbox{\quad\quad
    $_{I\!\! N}$}\vskip 1.4cm\hbox{\quad\quad $_{2I\!\! N}$}\vskip
    1.9cm\hbox{\eightrm dimension of the}\hbox{\eightrm
    field state-space}}

\setbox26=\vbox{\hbox{$\ket{2,0}$}\vskip 0.25cm\hbox{$c^*(g_{i_1})\phi_{1,1}^{(0)}$}\vskip
     1cm\hbox{$c^*(g_{i_1})(\phi_{0,1}^{(0)})_2$}\hbox{$c^*(g_{i_1})(\phi_{0,1}^{(0)})_1$}\vskip
     2cm\hbox{\eightrm
     \quad eigenvectors}}
\setbox27=\hbox{\box22\hskip 1.33cm\box23}
\setbox28=\vbox{\box21\vskip 0,83cm\box27}
\setbox29=\line{\box28\quad\box25\ \box00\quad\box24\quad\box26}

\vbox{\box29\medskip\hbox {\centerline{Figure 10. Levels of
    Figure 9, with eigenvectors of $H_0(\lambda)$ in ${\cal E}_2$;}}\hskip
    3cm\hbox{at most one spectator-photon}}

\footnote{}{$^{(1)}$ \eightrm dotted lines are slightly shifted from
  the level for reading purpose}

The degeneracy of the eigenvalue is greater if we admit eigenvectors
with a greater number of spectator-photons, with all their possible
states.

But considering a total number of photons greater than one yields other
eigenvalues. For instance, we claim that eigenvectors in ${\cal E}_{2}$ are not
necessarily of the form given in the last column of Figure 10. We are
going to see that the set of perturbed eigenvalues coming from one
given unperturbed energy $e_i$ changes when the number of photons
coupled to the atom changes. This was already the case for the
two-level atom of Section III (see for instance Proposition 3.1 of
Section III.2, for $\mu=0$, where the eigenvalue depended on the space
in which the eigenvector was looked for).

{\bf IV.2.3 Other eigenvectors of $H_0(\lambda)$ in ${\cal
  E}_{n+1}$ and other eigenvalues}

In this section, we assume $f_{01}\not=f_{12}$.

{\bf Proposition 4.3} {\it $H_0(\lambda)$ has an infinity of
eigenvalues different from $(\zeta_{0,1})_1,\break (\zeta_{0,1})_2$ et $\zeta_{1,1}$.} 

\underbar {Proof} ${\cal E}_n$ is invariant by $H_0(\lambda)$. The
subspace ${\cal E}_n^{(2)}$ of ${\cal E}_n$ is also
invariant.

\noindent 
Let us first consider ${\cal E}_2^{(2)}$, six-dimensional. If the
coupling constants are small, there are six eigenvalues: when
$\lambda_{01}$ and $\lambda_{12}$ tend to $0$, one of them tends to
$e_2$ (it is denoted by $\zeta_{2,2}$), two tend to $e_1$ (denoted by
$(\zeta_{1,2})_1$ and $(\zeta_{1,2})_2$) and three, one of which is
zero, tend to $0$ (denoted by $(\zeta_{0,2})_1=0$, $(\zeta_{0,2})_2$ and
$(\zeta_{0,2})_3$). They are obtained through diagonalizing a six by
six matrix (see Appendix B2). The calculation is straightforward
although the result has not a simple expression. In this 2-excitation
space, that two eigenvalues tend to $e_1$ is due to the fact that
there are two possible photon states, and that three eigenvalues tend
to $0$ is due to the fact that there are  three possible independent
states for the two photons. The fourth column
of Figure 11 recall these numbers. The eigenvector associated with
$(\zeta_{0,2})_1=0$ is $\ket{0,g_0\vee g_0}$; we met it under
the form $(\phi_{0,1}^{(1,0,\cdots)})_2$ (see (4.11)).

\noindent
We already found other eigenvectors in the same space ${\cal
E}_2$. They were built from eigenvectors in ${\cal E}_1^{(2)}$. They
are for instance $(\phi_{0,1}^{(1,0,\cdots)})_1$ and
  $\phi_{1,1}^{(1,0,\cdots)}$, with notations of Proposition
  4.2; they are in ${\cal E}_2^{(3)}$ if $g_1\in{\cal H}_{\rm
  rad}^{(3)}$. The corresponding eigenvalues,
  $(\zeta_{0,1})_1$ and $\zeta_{1,1}$, are different from the six we
just saw (see Appendix B2), except possibly for particular values of
$\lambda$.

Since the notations are a bit heavy, we again represent the levels in
Figure 11.

\setbox31=\psboxto(4.5cm;0cm){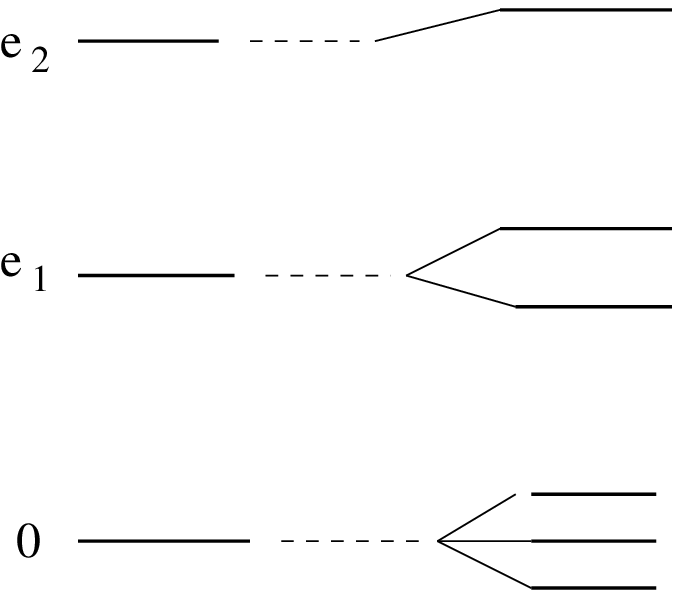}

\setbox32=\vbox{\hbox{\eightrm levels of the}\hbox{\eightrm decoupled}\hbox{\eightrm atom}}

\setbox33=\vbox{\hbox{\eightrm 2-excitation}\hbox{\eightrm mixed levels}}

\setbox34=\vbox{\hbox{\quad\quad 1}\vskip 1.35cm\hbox{\quad\quad 2}\vskip 1.4cm\hbox{\quad\quad
    3}\vskip 0.85cm\hbox{\eightrm dimension of the}\hbox{\eightrm
    field state-space}\hbox{\eightrm (apart from deg.)}}

\setbox35=\vbox{\hbox{$\zeta_{2,2}$}\vskip1.1cm\hbox{$(\zeta_{1,2})_2$}\hbox{$(\zeta_{1,2})_1$}\vskip
    0.82cm\hbox{$(\zeta_{0,2})_3$}\hbox{$(\zeta_{0,2})_1$}\hbox{$(\zeta_{0,2})_2$}\vskip0.89cm\hbox{\eightrm
    eigen-}\hbox{\eightrm values}}

\setbox36=\vbox{
\hbox{$\phi_{2,2}^{(0)}$\quad\raise 1mm\hbox{$_{\ket{2,0},\  \ket{1,1},\ 
      \ket{0,2}}$}}\vskip
0.8cm\hbox{$(\phi_{1,2}^{(0)})_2$\hskip 1cm ''}\hbox{$(\phi_{1,2}^{(0)})_1$\hskip 1cm ''}\vskip
0.6cm\hbox{$(\phi_{0,2}^{(0)})_3$\hskip 1cm ''}\hbox{$(\phi_{0,2}^{(0)})_1$\hskip 1cm ''}
\hbox{$(\phi_{0,2}^{(0)})_2$\hskip 1cm ''}\vskip 0.66cm      \hbox{\ \
\eightrm eigenvectors }\hbox{\ \ \eightrm and their contents  }}


\setbox37=\hbox{\box32\hskip 1.5cm\box33}

\setbox38=\vbox{\box31\vskip 0,83cm\box37}

\setbox39=\line{\box38\quad\box35\quad\box34\quad\box36}

\vbox{\box39\medskip\hbox{\centerline{Figure 11. Levels associated with
    eigenvectors of $H_0(\lambda)$ in ${\cal
    E}_2^{(2)}$,}}\hbox{\hskip 3.7cm without any spectator-photon}}

\medskip
\noindent
Figures in the fourth column are the dimensions of the projections of ${\cal E}_2^{(2)}$ on $\ket i\otimes {\cal F}^{(2)}$.

The changes in the levels $e_0$, $e_1$ and $e_2$ of Figures 10 and 11
cannot be superimposed on one another, in general. Both are to be considered in describing the levels of $H_0(\lambda)$.

To complete the study, we have to take an arbitrary number of
photons into account. The study of the spectrum of $H_0(\lambda)$ is
completed when one has also considered eigenvectors in ${\cal
E}_3^{(2)}$, ${\cal E}_4^{(2)}$, etc..; They give new
levels. Eventually, there is a very great number of levels. Let
$S(n,p)$ be the number of independent symmetric states that can be
formed with $n$ photons, each photon being in $p$ possible states. Let
us consider $1, 2, \cdots, n$ photons successively. The fundamental
state is split into a doublet, a triplet, $\cdots$, a
$S(n,2)$-multiplet. If we look for all possible states, all these
levels must be considered. As regards level 1, a shift, then a doublet,
a triplet, $\cdots$, a
$S(n-1,2)$-multiplet. Lastly, for level 2, we get no change, then a
shift, a doublet,  $\cdots$, a $S(n-1,2)$-multiplet. To be complete,
let us recall level $e_0=0$, with eigenvector $\ket{0,\Omega}$. \cqfd

The doubly infinite degeneracy due to spectator-photons still
remains. With the six eigenvalues $\zeta_{2,2}$,
$(\zeta_{1,2})_1$ and  $(\zeta_{1,2})_2$, and lastly  $(\zeta_{0,2})_1$, $(\zeta_{0,2})_2$ et
$(\zeta_{0,2})_3$, are associated eigenvectors $\phi_{2,2}^{(0)}$,
$(\phi_{1,2}^{(0)})_1$ and  $(\phi_{1,2}^{(0)})_2$), and lastly  $(\phi_{0,2}^{(0)})_1$, $(\phi_{0,2}^{(0)})_2$ et
$(\phi_{0,2}^{(0)})_3$.
As previously, through application of $\big(1\otimes c^*(g)\big)^n$ with $g$
orthogonal to $\{f_{01},f_{12}\}$, or more generally application of $\prod
\big(1\otimes c^*(g_i)\big)^{n_i}$, one gets other eigenvectors associated with
the same eigenvalues. In the same way that we built Figure 10 from Figure
9, we could illustrate this degeneracy graphically through adding
dotted lines in Figure 11.

Let us come back to our initial problem, which is to determine the
spectrum of $H(\lambda)$, at least roughly. It may be expected that
the perturbation $\lambda_{02} \ V$ partially removes the degeneracy
of each of the six above mentioned eigenvalues, as well as
degeneracies of the same type, for instance those of the three
eigenvalues $ (\zeta_{0,1})_1$,   $(\zeta_{0,1})_2$ and $ \zeta_{1,1}$
that we obtained previously. It is this simpler question that we now
 examine. We are going to show that the
coupling $\lambda_{02} \ V$ splits the first three levels of Figure
10, eigenvalues $ (\zeta_{0,1})_1$,   $
(\zeta_{0,1})_2$ and $ \zeta_{1,1}$ of Proposition 4.2, into an
infinity of levels and calculate the splittings of $ (\zeta_{0,1})_1$ and $ \zeta_{1,1}$,
at the lowest order in $\lambda_{02}$.  

\medskip
{\bf IV.3 Perturbation with respect to $\lambda_{02}$. Second removal
of degeneracy }

We are now interested in $H(\lambda)$. The function $f_{02}$ comes into
play. Hence we assume that the first vector
of the basis $g_1,g_2,\cdots$ of Proposition 4.2 is in ${\cal H}_{\rm
rad}^{(3)}$. The $g_i$'s, $i>1$ are thus orthogonal to ${\cal H}_{\rm
rad}^{(3)}$. We saw that the $\big(\phi
_{i,p}^{(q,0,\cdots)}\big)_k$, $q=0,1,\cdots$, which are in ${\cal
E}^{(3)}$, are associated with a unique eigenvalue $(\zeta_{i,p})_k$. In
the simple cases $(i,p)=(1,1)$ and $(i,p,k)=(0,1,1)$, we
are going to show that the degeneracy is removed. We calculate the
approximations of order two in $\lambda_{02}$ of those eigenvalues of $H(\lambda)$
which tend to $(\zeta_{i,p})_k$ when $\lambda_{02}$ tends to
$0$. These eigenvalues depend on $q$. The approximations are denoted
by $(z_{i,p}^{(q)})_k^{(\leq 2)}$. The corresponding eigenvectors are
denoted by $(\chi_{i,p}^{(q)})_k$ and their 2-order approximations by 
$(\chi_{i,p}^{(q)})_k^{(\leq 2)}$. We will have
$$
\lim_{\lambda_{02}\rightarrow
  0}(\chi_{i,p}^{(q)})_k^{(\leq 2)}=(\phi_{i,p}^{(q,0,\cdots)})_k=\big(1\otimes c^*(g_1)\big)^q(\phi_{i,p}^{(0)})_k   \eqno (4.13)
$$
(Here, $z$ has not the same meaning as in Section III; $\mu$ remains
zero.)

Note that, if $\chi\in{\cal H}^{(3)}$ is an eigenvector of
$H(\lambda)$, then for every $g$ orthogonal to ${\cal H}_{\rm
  rad}^{(3)}$, $\bigg(1\otimes\big(c^*(g)\big)^n\bigg)\chi$ is still
an eigenvector, associated with the same eigenvalue. This second
degeneracy removal is thus only very partial.

We assume that $p=1$ and we limit ourselves to perturbing $ (\zeta_{0,1})_1$,   $
(\zeta_{0,1})_2$ if $f_{01}\not=f_{12}$, and $ \zeta_{1,1}$.

\noindent 
If$f_{01}\not=f_{12}$, the diagram in Figure 10 transforms into the
one  of Figure 12.

\medskip
\setbox41=\psboxto(5.8cm;0cm){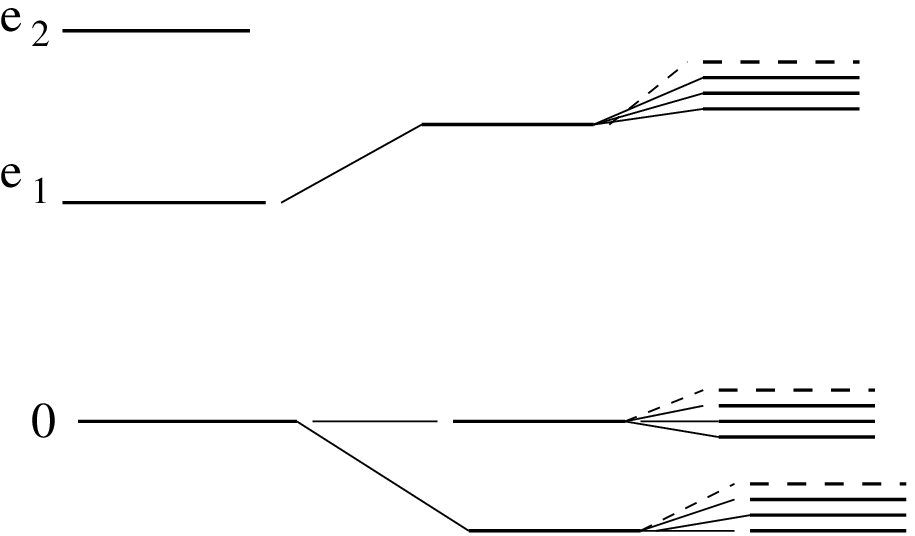}

\setbox42=\vbox{\hbox{\eightrm levels of the}\hbox{\eightrm decoupled}\hbox{\eightrm
atom}}

\setbox43=\vbox{\hbox{\eightrm levels}\hbox{\eightrm of \raise
    0.7mm\hbox {$_{H_0(\lambda)}$}}}

\setbox44=\vbox{\hbox{\eightrm levels}\hbox{\eightrm of \raise
    0.7mm\hbox {$_{H(\lambda)}$}}}

\setbox45=\vbox{\hbox{\quad\quad $_{I\!\!N}$}\vskip 1.6cm\hbox{\quad\quad
    $_{{I\!\!N}}^{\ \ 2}$}\vskip 0.32cm\hbox{\quad\quad
    $_{I\!\!N}$}\vskip 0.7cm\hbox{\eightrm dimension of the}\hbox{\eightrm
    field state-space}\hbox{\eightrm (apart from deg.)}}

\setbox46=\vbox{\hbox{$z_{1,1}^{(q)}$}\vskip 1.4cm\hbox{$(z_{0,1}^{(p,q)})_2$}\vskip
    0.2cm \hbox{$(z_{0,1}^{(q)})_1$}\hbox{$_{q=0,1,\cdots}$}\vskip 0.8cm\hbox{\eightrm
    eigen-}\hbox{\eightrm values}}

\setbox47=\vbox{
  \hbox
  {\hbox{$\chi_{1,1}^{(q)}$}\quad\vbox{\hbox{$\ \ _{\ket{0,1},\ket{0,2},..}$}\hbox{$\ \ _{\ket{1,0},\ket{1,1},..}$}}
     }\vskip1.4cm\hbox{$(\chi_{0,1}^{(p,q)})_2$\hskip 0.7 cm''}\vskip
  0.2cm\hbox{$(\chi_{0,1}^{(q)})_1$\hskip 0.9cm''}\hbox{$_{q=0,1,\cdots}$}\vskip 0.65cm      \hbox{\eightrm eigenvectors$^{(1)}$ and}\hbox{\eightrm
 their content}}

\setbox48=\hbox{\box42\hskip1.6cm\box43\quad\box44}

\setbox49=\vbox{\box41\vskip 0,83cm\box48}

\setbox50=\line{\box49\quad\box46\quad\box45\quad\box47}

\vbox{\box50\medskip\hbox {Figure 12.  Qualitative description
    of the perturbation of Figure 10 first levels, }\hskip
    0.9cm\hbox{ due to the $\lambda_{02}\  V$ term; ($g_{i_1}=g_1$)}}

\footnote{}{$^{(1)}$ \eightrm The residual degeneracy is not mentioned.}

\vskip 1cm
The case of $(\zeta_{0,1})_2$ is particular as we will see later on.

Note that the exact eigenvector $(\chi_{i,p}^{(q)})_k$ is no longer in 
${\cal E}_{q+1}^{(3)}$ but in $\displaystyle{\oplus_{r=0}^\infty {\cal
  E}_{r+1}^{(3)}}$. 

More precisely, we have 

{\bf Proposition 4.4}
{\it Through the
  perturbation $\lambda_{02}\ V$, each of the eigenvalues\break
  $(\zeta_{0,1})_1(\lambda_{01},\lambda_{12})$ and
  $\zeta_{1,1}(\lambda_{01},\lambda_{12})$ is at least split into an
infinity of eigenvalues\break $
(z_{0,1}^{(n)})_1(\lambda)$ and
$
z_{1,1}^{(n)}(\lambda) 
$, given at second order in $\lambda_{02}$ by the following formulas

(i)\hfill $ (z_{0,1}^{(n)})_1^{\leq
2}(\lambda)=(\zeta_{0,1})_1(\lambda_{01},\lambda_{12})+\lambda_{02}^2
\ \Big(1+n!\
{\lambda_{01}^2\over(\zeta_{0,1})_1^2}\Big)^{-1}\ A_{0,1,1}^{(n)}(\lambda_{01},\lambda_{12})$\hfill  {\rm
                                                (4.14)}  

\noindent
with, if $f_{01}\not=f_{12}$,
$$
A_{0,1,1}^{(n)}(\lambda_{01},\lambda_{12}):=n 
{\lambda_{01}^4\ |(f_{02},g_1)|^2 \over
\big((\zeta_{0,1})_1\big)^2\Big(\lambda_{12}^2(\zeta_{0,1})_1+\lambda_{01}^2\big((\zeta_{0,1})_1-e_2\big)\Big)}
                                            \eqno (4.15)
$$
where $|(f_{02},g_1)|$ can be expressed with the $s_i$'s,

\noindent
and, if $f_{01}=f_{12}$, and thus $g_1=(1+|s_0|^2)^{-{1\over2}}(f_{02}-s_0f_{01})$,
$$
A_{0,1,1}^{(n)}(\lambda_{01},\lambda_{12}):={\lambda_{01}^4\  \over
\big((\zeta_{0,1})_1\big)^2\Big(\lambda_{12}^2(\zeta_{0,1})_1+\lambda_{01}^2\big((\zeta_{0,1})_1-e_2\big)\Big)}(n|(f_{02},g_1)|^2+|s_0|^2)
                                             \eqno (4.16)
$$
\smallskip

(ii) $(z_{1,1}^{(n)})^{\leq 2}$ is obtained through replacing
$(\zeta_{0,1})_1$ by $\zeta_{1,1}$
in expressions giving $(z_{0,1}^{(n)})_1^{\leq 2}$.

(iii) $(\zeta_{0,1})_2$, which is more degenerated than the two
previous eigenvalues, is also split. The {\rm 2}-order approximations of
the perturbed eigenvalues, $(z_{0,1}^{(n,m})_2^{\leq 2}$, now
depending on two indices, are  obtained through the vanishing of an
infinite order determinant.}

The broad lines of the proof are given in Appendix B3, together with
the method for calculating the  corresponding eigenvectors $
(\chi_{0,1}^{(n)})_1^{\leq 2}(\lambda),\ 
(\chi_{0,1}^{(n,m)})_2^{\leq 2}(\lambda),\ $
 and $ 
(\chi_{1,1}^{(n)})^{\leq 2}(\lambda)
$

\medskip
{\bf IV.4 Conclusion of Section IV}

Section IV.3 described the splitting of the eigenvalues $
(\zeta_{0,1})_1$,   $(\zeta_{0,1})_2$ and $ \zeta_{1,1}$ of
Proposition 4.2. It gave a small part of the spectrum of
$H(\lambda,0)$, described in the second column of Figure 12. In view
of these results one may reasonably surmise the following points.

This splitting of the levels of the second column of Figure 9 (or 10)
will reproduce for those of the second column of Figure 11, which are different.
In other terms, the degeneracy of the latter levels, due to
spectator-photons, will also be removed by the coupling
$\lambda_{02}V$. More generally, each level multiplet which was
mentioned at the end of the proof of Proposition 4.3 is also split
when the interaction is totally switched on. This rough description of
the two level splittings we get by successively taking the two parts of the
interaction Hamiltonian into account eventually yields quite a
complicated spectrum for $H(\lambda,0)$. But the reason of the great
number of levels is simple; it is recalled in Section V below.

When $\mu$ increases from $0$, we expect that the eigenvalues we found
move into eigenvalues or resonances of $H(\lambda,\mu)$. The set of these
eigenvalues or resonances thus likely has the same rich structure.

\medskip
{\bf V General conclusion }

Results of Sections III and IV fit into the same frame. 
They lead us to
expect that the number of eigenstates or resonances of the atom-photon
system is formally, apart from accidental degeneracy, the product of the
dimensions of the atom state-space and the field state-space. The
description is complicated because of the multivaluedness of the
resolvent matrix elements, as functions of the $\lambda$ and $\mu$
parameters. 

\noindent
This picture may be illustrated through the following argument: 
if the energy of each level of the isolated atom is
considered as an eigenvalue of $H_{\rm atom}\otimes 1_{\rm field}$,
this level is twice infinitely degenerated (number of photons and
state of each photon). This degeneracy is removed with the actual
Hamiltonian which describes the coupling of the atom to the field. 
The shift of the naked-atom levels by the coupling of the atom to the
photon field is thus not the main feature in the change in the Hamiltonian
``spectrum''. The main feature is more the emergence of numerous resonances, as in the monochromatic-photon case.

For a two-level atom and Hamiltonian (3.1) (Section III), only one
photon state comes into play. The subspace of ${\cal H}_{\rm rad}$ to
be considered is ${\cal H}_{\rm rad}^{(1)}$ and the two degeneracies
which are removed, same energy for states $\ket{0,0},\ \ket{0,g},\cdots, \ket{0,g^{\vee
  n}},\cdots$ on the one hand and  $\ket{1,0},\ \ket{1,g},\cdots, \ket{1,g^{\vee
  n}},\cdots$ on the other hand, only concern the number of
photons. A great degeneracy remains since adding photons in states orthogonal to the distinguished
$g$-state do not change the energies.

For a three-level atom, the number of coupling functions in the
Hamiltonian is greater and this forced us to start to pay attention
to different
photon states. As a consequence, the just mentioned degeneracy now starts being
removed. 

For a real atom, with its infinity of levels, the splitting will still
be greater. Our perturbative treatment illustrates how
the different photon states may be taken into account
successively. Calculations will of course be impossible if some 
physically justified simplifications are not made.

The present limits of the study are the following.

Section III described resonances for a realistic Hamiltonian
($\mu\not=0$), but for a system with only two levels. However, even in
that simplified case, we are far from having found all the resonances
since we considered only one-or-two-excitation spaces, ${\cal E}_1$ or
${\cal E}_2$. It would be necessary to take more than two photons into
account. But the resonances are then given by more and more
complicated equations.

In Section IV, to be able to present a qualitative description of
resonances in a three-level system, we had to work in the
limit $\mu=0$. This first stage seems unavoidable to us if one wants to
solve the question completely. Doing this, we were able to take an
infinity of photons into account. But the calculations are only carried
to the second order in $\lambda_{02}$ and also the displacement of the
resonances when $\mu$ becomes non-zero is just qualitatively
mentioned.

However, we have seen that these partial results give new information about
hybrid states which are present in matter-field interactions such as
the interaction to which we borrowed our terminology: the interaction of atoms
(or molecules) with the electromagnetic field. We hope that concrete
problems will justify approximations making calculations possible.

\vskip 0.8cm

{\bf Appendix A. Estimation of corrective terms in the
Fredholm expansion of Proposition 3.4}

\smallskip 

{\bf Proposition}
{\it Set 
$
\varphi_2(\mu,z,p,q):=\mu^2\ D_2(\mu,z,p,q)
$
and
$$
\varphi_3(\mu,z,p,q):={|g(p)|\ |g(q)|\ (1+|p|)^{1/4}\ (1+|q|)^{1/4}\ (|p|-|q|)^2\over
  |f(z-\mu |p|)|^{1/2}\ 
 |f(z-\mu
 |q|)|^{1/2}\ |{z\over\mu}-2|p||^{1/2}\ |{z\over\mu}-2|q||^{1/2}\ |{z\over\mu}-|p|-|q||^2}
$$
$$
\varphi_4(\mu,z,p,q):={|g(p)|^{2/3}\ |g(q)|^{2/3}\ (1+|p|)^{1/4}\ (1+|q|)^{1/4}\ (|p|-|q|)^2\over
  |f(z-\mu |p|)|^{1/3}\ 
 |f(z-\mu
 |q|)|^{1/3}\ |{z\over\mu}-2|p||^{1/3}\ |{z\over\mu}-2|q||^{1/3}\ |{z\over\mu}-|p|-|q||^2}
$$
defined for $z<0$.
we have
$$
C_2(\lambda,\mu,z)=(\lambda^2/\mu)^2\ ||\varphi_2(\mu,z,.,.)||_1          \eqno(A1)
$$
$$
|C_3(\lambda,\mu,z)|<M_3(\lambda,\mu,z):=4\  (\lambda^2/\mu)^3\ ||\varphi_3(\mu,z,.,.)||_3^3\eqno(A2)
$$
Assuming that {\rm(3.19)} also holds for $n=4$, we also have
$$
|C_4(\lambda,\mu,z)|<M_4(\lambda,\mu,z):=4^2\  (\lambda^2/\mu)^4\ ||\varphi_4(\mu,z,.,.)||_6^6\eqno(A3)
$$}

\underbar{Proof} Use (3.19) and  H\"older's inequalities. \cqfd

We omit $\lambda$, which is fixed to $0.1$. Table 2 gives values for $C_2\big(\mu,z_{0,2}^1(\mu)\big)$,\break$
  M_3\big(\mu,z_{0,2}^1(\mu)\big)$, $M_4\big(\mu,z_{0,2}^1(\mu) \big)$
  and $\partial_z C_1(\mu,z^{(2)}_a(\mu))$, from
$$\displaylines{
\partial_z C_1(\mu,z)=-\lambda^2\int{|g(p)|^2\over
  f(\mu,z-\mu|p|)(z-2\mu|p|)}\Big({1\over  f(\mu,z-\mu|p|)}-{1\over
  z-2\mu|p|}\Big)\
dp-\hfill\cr\hfill\lambda^2\int\int{|g(p)|^2|g(q)|^2\ dp\ dq\over
  f^2(\mu,z-\mu|p|)(z-2\mu|p|)(z-\mu(p+q))^2}
}$$
By definition, $C_1\big(\mu,z_{0,1}^1(\mu)\big)=1$. We see how the terms of
the Fredholm expansion decrease with the order.

For $\mu=\mu_c$, $D(\mu_c,0)$ is close to $0$. Let us recall that we
have $f\big(\lambda,\mu_c(\lambda),0\big)=0$. It would be interesting
to see whether $D(\mu_c,0)$ vanishes or not. To try and answer this
question, let us estimate the error we made in calculating the zero
of $D(\mu,.)$ by the truncated series. Let us consider the following
expansion of $D(\mu,.)$ near $z_{0,2}^1(\mu)$
$$\displaylines{
1- C_1\big(\mu,z_{0,2}^1(\mu)\big)-\big(z-z_{0,2}^1(\mu)\big)\
\partial_zC_1\big(\mu,z_{0,2}^1(\mu)\big)+{1\over 2}
C_2\big(\mu,z_{0,2}^1(\mu)\big)+\hfill\cr\hfill{1\over
2}\big(z-z_{0,2}^1(\mu)\big)\
\partial_zC_2\big(\mu,z_{0,2}^1(\mu)\big)-{1\over 6}\Big(
C_3\big(\mu,z_{0,2}^1(\mu)\big)+\big(z-z_{0,2}^1(\mu)\big)\ \partial_zC_3\big(\mu,z_{0,2}^1(\mu)\big)\Big)
}$$
Let us assume that terms $(z-z_{0,2}^1(\mu))\partial_zC_2(\mu,z_{0,2}^1(\mu))$
and
$\big(z-z_{0,2}^1(\mu)\big)\partial_zC_3\big(\mu,z_{0,2}^1(\mu)\big)$
can be neglected. Then the correction to the zero is
$${1\over\partial_zC_1\big(\mu,z_{0,2}^1(\mu)\big)}\Big(     -{1\over
2}C_2\big(\mu,z_{0,2}^1(\mu)\big)+{1\over
6}C_3\big(\mu,z_{0,2}^1(\mu)\big)\Big)$$
whose principal term is of the order of $4\ 10^{-4}$. Since it is
precisely the order of $z_{0,2}^1(\mu_c)$, it is not possible to
answer the question.

\medskip

{\bf Appendix B Sketches of proofs of results in the 3-level case}

\smallskip
{\bf B1 Sketch of the proof of Proposition 4.2}.
 
Let $\Phi:=\ket{2,F_n}+\ket{1,F_{n+1}}+\ket{0,F_{n+2}}$ be an
eigenvector of $H_0(\lambda)$ in ${\cal E}_{n+2}$ associated with $z$,
one of the two eigenvalues $(\zeta_{0,1})_1$ or $\zeta_{1,1}$. It can
be shown that
$$
(z-e_2)^{-1}\lambda_{12}^2(f_{12},F_{n+1})\vee
 f_{12}=z^{-1}\ \lambda_{01}^2(f_{01},F_{n+1})\vee f_{01}\eqno ({\rm B}1.1)
$$
where 
$$
(f_{12},F_{n+1})(p_1,p_2,\cdots,p_n)=\int \overline {f}_{12}(p)\
F_{n+1}(p,p_1,p_2,\cdots,p_n)\ dp
$$
$$(f^{\vee n}\vee
h)(p_1,p_2,\cdots,p_{n+1})={1\over
  n+1}\sum_{i=1}^{n+1}\Big(\build\Pi_{j\not=i}^{n+1}f(p_j)\Big)\ h(p_i)
$$
From (B1.1), through decomposing $F_{n+1}$ on a basis of the
$(n+1)$-photon space built with $f_{01},f_{12}$ and the $g_i$'s, we get that
$F_{n+1}$ is a sum of states
$\prod_{i=1}^\infty\big( c^*(g_i)\big)^{n_i}\Omega$ with $\sum_i
n_i=n+1$. From relations expressing that $\Phi$ is an eigenvector, we
derive 
$F_n=0$ and, if $z\not=0$,
  $F_{n+2}=z^{-1}\sqrt{n+2}\ \lambda_{01}F_{n+1}\vee f_{01}$; this
  implies that $\Phi$ is in ${\cal G}_{0,1,1}$ if $z=(\zeta_{0,1})_1$
or in ${\cal
G}_{1,1}$ if $z=\zeta_{1,1}$. Hence (i) and (iii).

\noindent
If $z=(\zeta_{0,1})_2=0$, then $\Phi$ is an eigenvector if and only if
$F_{n}=0,\ F_{n+1}=0$ and $(f_{01},F_{n+2})=0$. Hence (ii).

\vfill\eject

{\bf B2 The six eigenvalues $\zeta_{2,2}$, $(\zeta_{1,2})_1$,
$(\zeta_{1,2})_2$, $(\zeta_{0,2})_1$, $(\zeta_{0,2})_2$ and
$(\zeta_{0,2})_3$, with eigenvectors in ${\cal E}_2^{(2)}$}.

The space spanned by vectors $$
\ket{2,\Omega},\ \ \ket{1,f_{12}},\ \ \ket{1,f_{01}},\
\ket{0,f_{12}\vee f_{01}},\ \ket{0,f_{01}\vee f_{01}}
$$
is invariant. Through adding $\ket{0,f_{12}\vee f_{12}}$, we get a
basis of ${\cal E}_2^{(2)}$, in which the matrix of $H_0$ is 
$$\pmatrix{e_2&\lambda_{12}&s_1\lambda_{12}&0&0&0\cr
\lambda_{12}&e_1&0&{\lambda_{01}\over\sqrt 2}&0&\sqrt 2 s_1
\lambda_{01}\cr
0&0&e_1&s_1{\lambda_{01}\over\sqrt 2}&\sqrt 2\lambda_{01}&0\cr
0&\sqrt 2\lambda_{01}&0&0&0&0\cr
0&0&\sqrt 2\lambda_{01}&0&0&0\cr
0&0&0&0&0&0\cr
}$$
One of the eigenvalues is $0$ and the eigenvector is the one mentioned
in the text. The other eigenvalues are the $\zeta$'s for which
$$
\zeta(\zeta-e_1)\Big(\zeta(\zeta-e_1)(\zeta-e_2)-3\lambda_{01}^2(\zeta-e_2)-\lambda_{12}^2\zeta\Big)+\zeta\lambda_{01}^2\big(2\lambda_{01}^2+(2-|s_1|^2)\lambda_{12}^2\big)-2e_2\lambda_{01}^4
$$
vanishes. Neglecting fourth order terms, we get the following
solutions at second order in $\lambda$
$$\displaylines{
(\zeta_{0,2}^0)_2=0,\quad (\zeta_{0,2}^0)_3=-{3\over e_1}\lambda_{01}^2,\quad (\zeta_{1,2}^0)_1=e_1,\quad
(\zeta_{1,2}^0)_2=e_1+{3\over e_1}\lambda_{01}^2+{1\over
e_1-e_2}\lambda_{12}^2,\cr \zeta_{2,2}^0=e_2+{1\over
e_2-e_1}\lambda_{12}^2
}$$
This proves that these eigenvalues are different from those found in  ${\cal E}_1^{(2)}$.

\medskip

{\bf B3 Sketch of the proof of Proposition 4.4}.

Let us use Kato's method$^{23}$ to determine the three (infinite) sets
of perturbed eigenvalues which tend to each of the unperturbed
eigenvalues. (See a short account in Ref. 24.)

(i) Let us first consider $(\zeta_{0,1})_1$. The unperturbed
eigenspace is  ${\cal G}_{0,1,1}$ (Proposition 4.2). Let $P_0^{0,1,1}$
be the projector on this space and $Q_0^{0,1,1}=1-P_0^{0,1,1}$. We
need operator $\tilde Q_0^{0,1,1}$ which is sometimes written as $Q_0^{0,1,1}[(\zeta_{0,1})_1-H_0]^{-1}Q_0^{0,1,1}$; for
$x\in{\cal E}$, it is defined by $\tilde Q_0^{0,1,1}\ x= Q_0^{0,1,1}\
z$, where $z$ is any vector in ${\cal E}$ satisfying
$[(\zeta_{0,1})_1-H_0]\ z=Q_0^{0,1,1}\ x$. Let  ${\cal H}_{0,1,1}$ be
the direct sum of the eigenspaces associated with eigenvalues of
$H(\lambda)$ which tend to $(\zeta_{0,1})_1$ when $\lambda_{02}$ tends
to $0$. Let $P^{0,1,1}(\lambda_{02})$ be the projector
  on ${\cal H}_{0,1,1}$. Therefore
  $P^{0,1,1}(\lambda_{02})\build{\longrightarrow}_{\lambda_{02}\rightarrow 0}^{}
  P_0^{0,1,1}$.
We assume that $P_0^{0,1,1}$ and $P^{0,1,1}(\lambda_{02})$ establish
one-to-one correspondences  ${\cal H}_{0,1,1}\rightarrow{\cal
  G}_{0,1,1}$ and ${\cal
  G}_{0,1,1}\rightarrow{\cal H}_{0,1,1}$. Set
$$
L_{0,1,1}=P_0^{0,1,1}HP^{0,1,1}P_0^{0,1,1}\quad {\rm and}\quad
K_{0,1,1}=P_0^{0,1,1}P^{0,1,1}P_0^{0,1,1}             \eqno ({\rm B}3.1)
$$
We recall$^{24}$ the expansion of $P^{0,1,1}$: with
$S^{(0)}:=-P_0^{0,1,1}$ and $S^{(k)}:=(\tilde Q_0^{0,1,1})^k$,
$$
P^{0,1,1}=P_0^{0,1,1}-\sum_n\ \lambda_{02}^n\ \sum_{k_i\geq 0,\
  k_0+\cdots+k_n=n}\ S^{(k_0)}VS^{(k_1)}\cdots VS^{(k_n)}  \eqno ({\rm B}3.2)
$$
A necessary and sufficient condition for $\chi\in{\cal
  H}_{0,1,1}$ to be an eigenvector of $H(\lambda)$ associated with the
eigenvalue $z$ is that there exists $\phi\in{\cal G}_{0,1,1}$
satisfying $\chi=P^{0,1,1}\phi$ and
$$
L_{0,1,1}\ \phi=z\ K_{0,1,1}\ \phi                        \eqno ({\rm B}3.3)
$$
$L_{0,1,1}$ and $K_{0,1,1}$ are operators in ${\cal G}_{0,1,1}$ and
the problem of the perturbation of $(\zeta_{0,1})_1$ is turned into
finding such $z$. We still have a degeneracy due to photon-states in $\big({\cal H}_{\rm
rad}^{(3)}\big)^{\bot}$; indeed, if $\phi$ is a solution for (B3.3),
then $\Pi_{i\geq 2}\big(1\otimes c^*(g_i)\big)^{n_i}\phi$ is still a
solution. Thus we are not going to look for all $\phi$'s, but only for
those in an invariant subspace of ${\cal G}_{0,1,1}$. Lemmas B3.1
to B3.6 prepare the calculation of eigenvalues into which
$(\zeta_{0,1})_1$ splits. The result for $\zeta_{1,1}$ will be
obtained through a simple change in the notations. The splitting of
$(\zeta_{0,1})_2$ is just outlined.

Let $g_1$ be the function of
${\cal H}_{\rm rad}^{(3)}$ orthogonal to ${\cal H}_{\rm
  rad}^{(2)}$ and let ${\cal G}_{0,1,1}^{(1)}$ be the subspace of
${\cal G}_{0,1,1}$ spanned by
$\big(1\otimes c^*(g_1)\big)^n\ (\phi_{0,1}^{(0)})_1$, $n\geq 0$. Let us denote
 the approximations of  $K_{0,1,1}$ and $L_{0,1,1}$
to order $q$ in $\lambda_{02}$ by $K_{0,1,1}^{\leq q}$ and $L_{0,1,1}^{\leq q}$.

{\bf Lemma B3.1}
{\it For all $q$, ${\cal G}_{0,1,1}^{(1)}$ is invariant by $K_{0,1,1}^{\leq q}$
and $L_{0,1,1}^{\leq q}$.}

\smallskip\noindent
\underbar{Proof} Operators $1\otimes c^*(g_1)$ and $V$ sent ${\cal E}^{(3)}$
into ${\cal E}^{(3)}$, since functions orthogonal to ${\cal H}_{\rm
rad}^{(3)}$ do not play any part. Now, $P_0^{0,1,1}$ sends ${\cal E}^{(3)}$ into
${\cal G}_{0,1,1}^{(1)}\subset {\cal E}^{(3)} $. Thus, if $X$ is any
endomorphism of ${\cal E}^{(3)}$, $P_0^{0,1,1}XP_0^{0,1,1}$ is an
endomorphism of ${\cal G}_{0,1,1}^{(1)}$.
 $Q_0^{0,1,1}$ also leaves  ${\cal E}^{(3)}$
invariant. The same is true for $\tilde Q_0^{0,1,1}$ and therefore for
$P^{0,1,1}$. Hence the Lemma. \cqfd

Expressions to the second order in $\lambda_{02}$ of $L_{0,1,1}$ and
$K_{0,1,1}$ are
$$
K_{0,1,1}^{\leq 2}=P_0^{0,1,1}-\lambda_{02}^2  P_0^{0,1,1}\ V\  (\tilde
Q_0^{0,1,1})^2\ V\  P_0^{0,1,1}                           \eqno({\rm  B}3.4)
$$
$$
L_{0,1,1}^{\leq 2}=(\zeta_{0,1})_1 K_1^{\leq 2}+\lambda_{02}^2\  P_0^{0,1,1}\ V\ 
\tilde Q_0^{0,1,1}\ 
V\ P_0^{0,1,1}                   \eqno({\rm  B}3.5)
$$

\noindent
To calculate $K_{0,1,1}^{\leq 2}(\phi_{0,n+1}^{(n)})_1$ and $L_{0,1,1}^{\leq
  2}(\phi_{0,n+1}^{(n)})_1$, we need the following lemma:

\smallskip
{\bf Lemma B3.2}

 {\it  
$$
V\ (\phi_{0,n+1}^{(n)})_1=\ket{2,(\varphi_{0,1}^{(n)})_1} \eqno({\rm  B}3.6)
$$
with $(\varphi_{0,1}^{(0)})_1=
N_1\ {\lambda_{01}\over(\zeta_{0,1})_1}\overline s_0$ and,for $n\geq 1$,
$$
(\varphi_{0,1}^{(n)})_1=N_1\ {\lambda_{01}\over(\zeta_{0,1})_1}\
\big(\overline s_0\ g_1^{\vee^n}+ 
n(f_{02},g_1)\ f_{01}\vee g_1^{\vee^{n-1}}\big)           \eqno({\rm  B}3.7)
$$
where $N_1=(1+(\zeta_{0,1})_1^{-2}\lambda_{01}^2)^{-1/2}$}

\smallskip

{\bf Lemma B3.3} {\it $K_{0,1,1}^{\leq 2}$
and $L_{0,1,1}^{\leq 2}$ are diagonal in the basis
$(\phi_{0,n+1}^{(n)})_1$ of ${\cal G}_{0,1,1}^{(1)}$.} 

\smallskip\noindent
\underbar{Proof} $V(\phi_{0,n+1}^{(n)})_1\in{\cal E}_{n+2}$, since
$(\phi_{0,n+1}^{(0)})_1$ has no component on $\ket 2\otimes {\cal
F}$. Through using $\tilde Q_0^{0,1,1}{\cal E}_{n+2}^{(3)}\subset {\cal
  E}_{n+2}^{(3)}$, we then get 
$$
\big(\tilde Q_0^{0,1,1}\ V\
P_0^{0,1,1}(\phi_{0,n+1}^{(n)})_1,\tilde Q_0^{0,1,1}\ V\
P_0^{0,1,1}(\phi_{0,m+1}^{(m)})_1\big) =0\ ,\quad {\rm if}\quad
m\not=n
$$ and $K_{0,1,1}^{\leq 2}$ is
 diagonal. The same is true for $L_{0,1,1}^{\leq 2}$. \cqfd

{\bf Lemma B3.4} {\it Let us set $
u_{1,n}:=\tilde Q_0^{0,1,1}\ V\
(\phi_{0,n+1}^{(n)})_1
$ and define $\kappa_{1,n},\ \theta_{1,n+1}$ and $h_{1,n+2}$ by
$u_{1,n}=\ket{2,\kappa_{1,n}}+\ket{1,\theta_{1,n+1}}+\ket{0,h_{1,n+2}}$.
We denote the expression of $z$ to the second order in
$\lambda_{02}$ by $(z_{0,1}^{(n)})_1^{\leq 2}$. We have
$$(z_{0,1}^{(n)})_1^{\leq 2}=(\zeta_{0,1})_1+\lambda_{02}^2\ 
\big((\varphi_{0,1}^{(n)})_1,\kappa_{1,n}\big)        \eqno({\rm  B}3.8)
$$} 
\noindent
\underbar{Proof} From Lemma B3.3, it follows that
$(\phi_{0,n+1}^{(n)})_1$ are eigenvectors of $L_{0,1,1}^{\leq 2}$ and
$K_{0,1,1}^{\leq 2}$, associated with eigenvalues which we denote by $(l_{0}^{(n)})_1^{\leq 2}$ and
$(k_{0}^{(n)})_1^{\leq 2}$ respectively. Vectors $\phi$ in ${\cal
  G}_{0,1,1}^{(1)}$ satisfying $(L_{0,1,1}^{\leq
2}-zK_{0,1,1}^{\leq 2})\phi=0$ are necessarily these
$(\phi_{0,n+1}^{(n)})_1$; the corresponding $z$-value, for each $n$,
is $\big((k_{0,1}^{(n)})_1^{\leq 2}\big)^{-1}(l_{0,1}^{(n)})_1^{\leq 2}$.
Through using (B3.4) and (B3.6), we get
$$
(k_{0,1}^{(n)})_1^{\leq 2}=\big((\phi_{0,n+1}^{(n)})_1,K_{0,1,1}^{\leq
2}(\phi_{0,n+1}^{(n)})_1\big)=1-\lambda_{02}^2||u_{1,n}||^2  \eqno({\rm  B}3.9)
$$
$$
(l_{0,1}^{(n)})_1^{\leq
2}=(\zeta_{0,1})_1\Big(1-\lambda_{02}^2||u_{1,n}||^2\Big)+\lambda_{02}^2\big( 
V\ (\phi_{0,n+1}^{(n)})_1,u_{1,n}\big)   
$$
As a consequence,
$$
(l_{0,1}^{(n)})_1^{\leq
2}=(\zeta_{0,1})_1\Big(1-\lambda_{02}^2||u_{1,n}||^2\Big)+\lambda_{02}^2\big((\varphi_{0,1}^{(n)})_1,\kappa_{1,n}\big)                    \eqno({\rm  B}3.10)
$$
Hence (B3.8) holds since $1-||u_{1,n}||^2$ is to be replaced by 1, to
the considered approximation. \cqfd

{\bf Lemma B3.5} {\it 

(a) Let us set $M_1:=\Big({\lambda_{12}^2\over
  (\zeta_{0,1})_1-e_2}+{\lambda_{01}^2\over (\zeta_{0,1})_1}\Big)^{-1}$.
For $n\geq 0$, a vector $v$ satisfying\break
  $[(\zeta_{0,1})_1-H_0]v=\ket{2,(\varphi_{0,1}^{(n)})_1}$ is
  $v_{1,n}=\ket{2,\kappa'_{1,n}}+\ket{1,\theta'_{1,n+1}}+\ket{0,h'_{1,n+2}}$, with
$$
\kappa'_{1,n}={(\varphi_{0,1}^{(n)})_1\over(\zeta_{0,1})_1-e_2}+\sqrt{n+1}\
  \lambda_{12}{(f_{12},\theta'_{1,n+1})\over (\zeta_{0,1})_1-e_2}\eqno({\rm  B}3.11)
$$
$$
h'_{1,n+2}=\sqrt{n+2}\ \lambda_{01}{\theta'_{1,n+1}\vee f_{01}\over (\zeta_{0,1})_1}\
                   \eqno({\rm  B}3.12)
 $$
$$
\theta'_{1,n+1}=N_1\sqrt{n+1}{\lambda_{01}\lambda_{12}\over(
  (\zeta_{0,1})_1-e_2) (\zeta_{0,1})_1}\big(
s_0(\theta'_{1,n+1})^{(1)}+n(f_{02},g_1)(\theta'_{1,n+1})^{(2)}\big)
                                           \eqno({\rm  B}3.13)
$$
where, if $f_{01}\not=f_{12}$,
$$
(\theta'_{1,n+1})^{(1)}={(\zeta_{0,1})_1-e_2\over \lambda_{12}^2}{1\over
  1-|s_1|^2}\ (s_1f_{01}-f_{12})\vee g_1^{\vee^n}      \eqno({\rm  B}3.14)
$$
$$
(\theta'_{1,1})^{(2)}=0,\quad(\theta'_{1,n+1})^{(2)}={M_1\over
 2( 1-|s_1|^2)}\Big(s_1f_{01}\vee f_{01}-2f_{01}\vee f_{12})+\overline s_1f_{12}\vee f_{12}\Big)\vee
g_1^{\vee^{n-1}}                                   \eqno({\rm  B}3.15)
$$
If $f_{01}=f_{12}$,
$$
(\theta'_{1,n+1})^{(1)}=-M_1\ f_{01}\vee g_1^{\vee n} \eqno({\rm  B}3.16)
$$
$$
(\theta'_{1,1})^{(2)}=0,\quad(\theta'_{1,n+1})^{(2)}=-{1\over 2}M_1\ f_{01}\vee f_{01}\vee g_1^{\vee (n-1)} \eqno({\rm  B}3.17)
$$

(b) $u_{1,n}$ of  Lemma {\rm  B3.4} is equal to $v_{1,n}$}

\smallskip\noindent
\underbar{Proof} Relations  (B3.11) et (B3.12) are obtained through
projecting the equality defining $v$ on $\ket 2\otimes {\cal H}_{\rm rad}$ and
$\ket 0\otimes {\cal H}_{\rm rad}$, respectively. The projection on
$\ket 1\otimes {\cal H}_{\rm rad}$ implies that $\theta'_{1,n+1}$ satisfies
$$
{\cal L}_{1,n+1}\theta'_{1,n+1}={\sqrt{n+1}\lambda_{12}\over
  (\zeta_{0,1})_1-e_2}(\varphi_{0,1}^{(n)})_1\vee f_{12}\eqno ({\rm B}3.18)
$$
where
$$
{\cal L}_{1,n+1}\theta'_{1,n+1}:=-{(n+1)\lambda_{12}^2\over(\zeta_{0,1})_1-e_2}
f_{12}\vee(f_{12},\theta'_{1,n+1})-{(n+1)\lambda_{01}^2\over(\zeta_{0,1})_1}
f_{01}\vee(f_{01},\theta'_{1,n+1})                           \eqno ({\rm B}3.19)
$$
For $n\geq 1$, $(\varphi_{0,1}^{(n)})_1$, given by (B3.7) is
decomposed into two parts. Hence we introduce two functions $(\theta'_{1,n+1})^{(1)}$ and
$(\theta'_{1,n+1})^{(2)}$ satisfying
$$
{\cal L}_{1,n+1}(\theta'_{1,n+1})^{(1)}=g_1^{\vee n}\vee f_{12}  \eqno ({\rm B}3.20)
$$
$$
{\cal L}_{1,n+1}(\theta'_{1,n+1})^{(2)}=g_1^{\vee (n-1)}\vee f_{01}\vee f_{12}\eqno ({\rm B}3.21)
$$
so that a solution of (B3.18) will be given by (B3.13). To prove (a)
if $f_{01}\not=f_{12}$, we check that (B3.14) satisfies (B3.20) and that (B3.15) satisfies
(B3.21). We proceed in the same way with (B3.16) and (B3.17), if
$f_{01}=f_{12}$, (B3.18) and (B3.19) being still true. Only
$(\theta_{1,n+1})^{(1)}$ plays a part if $n=0$.

To prove (b), we note that vectors in ${\cal G}_{0,1,1}$ are linear
combinations of $\ket{1,g_{i_1}\vee\cdots g_{i_n}}$ and $\ket{0,f_{01}\vee
g_{i_1}\vee\cdots g_{i_n} }$. Since (a) implies that $\theta'_{1,n+1}$
is a sum of symmetric products of terms one of which at least is in ${\cal H}_{\rm
  rad}^{(2)}$, we have $(\theta'_{1,n+1},g_{i_1}\vee\cdots g_{i_{n+1}})=0$ and
$(h'_{1,n+2},f_{01}\vee g_{i_1}\vee\cdots g_{i_{n+1}})=0$. As a consequence, $\ket{2,\kappa'_{1,n}}$,
$\ket{1,\theta'_{1,n+1}}$ and $\ket{0,h'_{1,n+2}}$ are orthogonal to
${\cal G}_{0,1,1}$ and $u_{1,n}=Q_0^{0,1,1}v_{1,n}=v_{1,n}$. \cqfd

\smallskip

{\bf Lemma B3.6} {\it Take $n\geq 0$}. For $f_{01}\not=f_{12}$,
$$
\big((\varphi_{0,1}^{(n)})_1,\kappa_{1,n}\big)=n  N_1^2
{\lambda_{01}^4\ |(f_{02},g_1)|^2 \over \big((\zeta_{0,1})_1\big)^2\Big(\lambda_{12}^2(\zeta_{0,1})_1+\lambda_{01}^2\big((\zeta_{0,1})_1-e_2\big)\Big)}
\eqno ({\rm B}3.22)
$$
For $f_{01}=f_{12}$,
$$
\big((\varphi_{0,1}^{(n)})_1,\kappa_{1,n}\big)=  N_1^2
{\lambda_{01}^4\  \over \big((\zeta_{0,1})_1\big)^2\Big(\lambda_{12}^2(\zeta_{0,1})_1+\lambda_{01}^2\big((\zeta_{0,1})_1-e_2\big)\Big)}(n|(f_{02},g_1)|^2+|s_0|^2)
\eqno ({\rm B}3.22')
$$

\smallskip\noindent
\underbar{Proof} Use (B3.7) and Lemma B3.5. \cqfd

Formulas (4.14) and (4.15) then follow from (B3.10)
and (B3.22). In the same way, (4.16) follows from (B3.22').

\medskip
(ii) All that has been written up to now can be transposed from
$(\zeta_{0,1})_1$ to $\zeta_{1,1}$, with the following changes
$$\displaylines{
P_0^{0,1,1}\rightarrow P_0^{1,1},\ 
P^{0,1,1} \rightarrow P^{1,1},\ 
Q_0^{0,1,1} \rightarrow Q_0^{1,1},\ 
\tilde Q_0^{0,1,1} \rightarrow\tilde Q_0^{1,1},\ 
{\cal H}_{0,1,1}\rightarrow {\cal H}_{1,1},\cr
 (\phi_{0,n+1}^{(n)})_1\rightarrow  \phi_{1,n+1}^{(n)},\ 
  {\cal G}_{0,1,1}\rightarrow   {\cal G}_{1,1},\  
{\cal G}_{0,1,1}^{(1)}\rightarrow  {\cal G}_{1,1}^{(1)},\ 
(\varphi_{0,1}^{(n)})_1\rightarrow \varphi_{1,1}^{(n)},\cr 
 (k_{0,1}^{(n)})_1 \rightarrow k_{1,1}^{(n)},\ 
 (l_{0,1}^{(n)})_1\rightarrow l_{1,1}^{(n)},\ 
(z_{0,1}^{(n)})_1^{\leq 2}\rightarrow (z_{1,1}^{(n)})^{\leq 2}
}$$
$N_1$, $M_1$, $(\varphi_{0,1}^{(n)})_1$ are thus also changed, as well
as ${\cal L}_{1,n+1}$ and the solutions
$\theta_{1,n+1}$\break $h_{1,n+2}$ and $\kappa_{1,n}$. 

\noindent
One then get (ii) of  Proposition 4.4.

\medskip
(iii) Let us come on now to the splitting of $(\zeta_{0,1})_2$, an
eigenvalue which is zero and exists only if $f_{01}\not=f_{12}$. The
issue is more complicated due to the fact that the unperturbed
eigenspace of interest is no longer just spanned by the $\Pi (1\otimes c^*(g_1))^n
(\phi_{0,1}^{(0)})_2$. It is spanned by the $\ket{0,\Pi \big( c^*(g_i)\big)^{n_i}
\Omega}$, with $i=0\ {\rm or}\ 1$. Let us denote this space by ${\cal G}$.
Let $P_0^{0,1,2}$ be the projector on ${\cal G}$. With $K_{0,1,2}^{\leq
2}$ and $L_{0,1,2}^{\leq 2}$ defined as in (B3.4) and (B3.5), we can
see that $L_{0,1,2}^{\leq 2}$ has non-vanishing matrix elements between
say  $g_0^{\vee p}\vee g_1^{\vee (p+2)}$ and
$g_0^{\vee (p+1)}\vee g_1^{\vee (p+1)}$, or between $g_0^{\vee (p+1)}\vee g_1^{\vee (p+1)}$ and
$g_0^{\vee (p+2)}\vee g_1^{\vee (p)}$ or also between $g_0^{\vee (p+2)}\vee g_1^{\vee (p)}$ and
$g_0^{\vee (p+3)}\vee g_1^{\vee (p-1)}$. This makes the computation of the
$z$'s satisfying ${\rm det}(L_{0,1,2}^{\leq 2}-z\ K_{0,1,2}^{\leq
2})=0$ more intricate and we don't calculate them here.

This completes the proof of Proposition 4.4. \cqfd
\smallskip

Eigenvectors  $
(\chi_{0,1}^{(n)})_1^{(2)},\ 
(\chi_{1,1}^{(n)})^{(2)}
$
are obtained from the correspondences\break
${\cal G}_{0,1,1}\rightarrow{\cal H}_{0,1,1}$,
and
${\cal G}_{1,1}\rightarrow{\cal H}_{1,1}$, through the second order
expansion of operators
$P^{0,1,1}(\lambda_{02})$ and $ P^{1,1}(\lambda_{02})$ which perform these correspondences.
For example, we get (see Ref 24 p. 614)
$$
(\chi_{0,1}^{(n)})_1^{(2)}=(\phi_{0,n+1}^{(n)})_1+\lambda_{02}^2\Big(-P_0^{0,1,1}V(\tilde
Q_0^{0,1,1})^2V+\tilde Q_0^{0,1,1}V\tilde
Q_0^{0,1,1}V\Big)\ (\phi_{0,n+1}^{(n)})_1
$$

\vfill\eject
\centerline {\bf References}

\smallskip

$^{\ 1}$V. Bach, J. Fr{\"o}hlich, I.M. Sigal, Adv. Math. {\bf 137} 299
(1998)

$^{\ 2}$A. Arai, M. Hirokawa, F. Hiroshima, preprint,
arXiv:math-ph/0409055 (2004)

$^{\ 3}$M. Hirokawa, Physics Letters A {\bf 294} 13 (2002)

$^{\ 4}$M. Hirokawa, Rev. Math. Phys. {\bf 13} 221 (2001)

$^{\ 5}$M. Griesemer, E. H. Lieb, M. Loss, Invent. Math.  {\bf 145} 557 (2001)

$^{\ 6}$R. Minlos, H. Spohn, Amer. Math. Soc. Transl. {\bf 177} 159 (1996)

$^{\ 7}$M. H{\"u}bner, H. Spohn, Ann. Inst. Henri Poincar\'e {\bf 62} 289 (1995)

$^{\ 8}$A. Arai, J. Math. Anal. Appl. {\bf 140}, 270 (1989)

$^{\ 9}$C. Cohen-Tannoudji, J. Dupont-Roc, G. Grynberg,  
 Processus
d'Interac\-tion entre\break
\indent $\ \ $ Photons et Atomes (Inter\'editions/Editions du
C.N.R.S. Paris 1988 ; (English trans-\break
\indent $\ \ $ lation: J. Wiley, New York 1992).

$^{10}$C. Billionnet,  J. Phys. A {\bf 35}, 2649 (2002);
Int. J. Mod. Phys. A {\bf 19}, 2643 (2004)

$^{11}$B. Sermage, S. Long, I. Abram, J. Y. Marzin, J. Bloch,
R. Planel, V. Thierry-Mieg,
\indent $\ \ $ Phys. Rev. B {\bf 53}, 16516 (1996)

$^{12}$C. Weisbuch, M. Nishioka, A. Ishikawa, Y. Arakawa,  
    Phys. Rev. Lett. {\bf 69}, 3314\break 
\indent $\ \ $ (1992)

$^{13}$S. Hameau, Y. Guldner, O. Verzelen, R. Ferreira,
 G. Bastard, J. Zeman, A. Lema\^\i{}tre, 
\indent $\ \ $ J. M. G\'erard, Phys. Rev. Lett. {\bf 83}, 4152 (1999)

$^{14}$T. Inoshita, H. Sakaki,   Phys. Rev. B {\bf 56}, R4355 (1997)

$^{15}$J. Tignon, P. Voisin, C. Delalande, M. Voos, R. Houdr\'e,
U. Oesterle, R. P. Stanley, 
\indent $\ \ $ Phys. Rev. Lett. {\bf 74}, 3967 (1995)

$^{16}$S. Haroche, J. M. Raimond,  Pour la Science {\bf 188} (1993)

$^{17}$C. Billionnet,   Lett. Math. Phys. {\bf 54}, 61 (2000)

$^{18}$C. Billionnet,   J. Phys. A {\bf 34},  7757 (2001)

$^{19}$H. E. Moses,  Lett. Nuov. Cim. {\bf 4}, 51 (1972). 

$^{20}$K. O. Friedrichs,  Comm. Pure Appl. Math. {\bf 1},
361 (1948)

$^{21}$C. Billionnet,  Ann. H. Poincar{\'e}. {\bf 2}, 361 (2001).

$^{22}$F.G. Tricomi,   Integral equations (Dover
  Publications, New York 1985)

$^{23}$T. Kato,  Perturbation Theory for Linear Operators
(Springer, Berlin 1987)

$^{24}$A. Messiah,   M\'ecanique Quantique II, p. 614-615
(Dunod, Paris 1964)

\end